\documentclass[journal=jpccck,manuscript=article,layout=twocolumn]{achemso}

\usepackage[version=3]{mhchem} 
\usepackage{appendix}
\usepackage{amsmath}
\usepackage{mciteplus}
\usepackage{color}

\newcommand{\ep}{\varepsilon}
\author{Shumpei Masuda}
\affiliation
{James Franck Institute, The University of Chicago, Chicago, IL 60637, US}
\alsoaffiliation
{Department of Physics, Tohoku University, Sendai 980, Japan}
\author{Stuart A. Rice}
\affiliation
{James Franck Institute, The University of Chicago, Chicago, IL 60637, US}
\email{s-rice@uchicago.edu}

\title[An \textsf{achemso} demo]
  {Selective Vibrational Population Transfer using Combined STIRAP and Counter-Diabatic Fields}

\begin{document}







\begin{abstract}
We report studies of state-to-state vibrational energy transfer in an isolated polyatomic molecule driven by combined stimulated Raman adiabatic passage (STIRAP) and counter-diabatic fields (CDF), using as vehicles selective population of one of a pair of near degenerate states in SCCl$_2$ and enhancing the yield of HNC in the HCN$\rightarrow$HNC isomerization reaction.  The efficiency of the population transfer within a subset of states embedded in a dense manifold of states is examined for the cases that the transition dipole moments between the initial and target states are much smaller than and comparable to the transition dipole moments between background states, and for the case that the subset states have large transition dipole moments with the background states.  We show that, in a subset of states that is coupled to background states, a combination of STIRAP fields and CDFs that do not individually generate processes that are competitive with the desired population transfer can generate greater population transfer efficiency than can ordinary STIRAP with field strength and/or pulse duration similarly restricted.  We also show that the exact CDF for an isolated three level system is a useful approximation to the exact CDF for three and five state systems embedded in background states, and we report a study of the stability of the STIRAP + approximate CDF control protocol to variation of the CDF.
\end{abstract}

\section{Introduction}
In the last 25 years theoretical and experimental studies have established several methodologies that provide active control of the quantum dynamics of a system via manipulation of the frequency, phase and temporal character of an applied optical field \cite{Rice,Shapiro}.  The underlying mechanisms of all the methods rely on coherence and interference effects embedded in the quantum dynamics; they vary in efficiency, generality of application, and sensitivity to perturbations.  The perturbations referred to can arise from interactions between a system and its surroundings and/or from dynamical effects associated with neglected terms in a simplified Hamiltonian used to characterize a system.  In general, it is not possible to calculate the exact control field driving a process in an $N$-level system, and it is likely that if calculated that field is too complicated to be realized experimentally.  For those reasons, often a control methodology is developed with respect to field-generated manipulation of population within a subset of states, without regard for the influence of the remaining background states on the quantum dynamics.  In this paper we address this issue.  Specifically, we examine the influence of a dense manifold of background states on population transfer within a subset of embedded states, and on how that influence can be suppressed.  The vehicle for our study is a particular active control protocol, namely, driving selective vibrational energy transfer in a polyatomic molecule by an assisted adiabatic process.  Demirplak and Rice developed the counter-diabatic control protocol while studying control methods that efficiently transfer population between a selected initial state and a selected target state of an isolated molecule \cite{Demi,Demi2,Demi3}. The protocol has been studied for manipulation of atomic and molecular states \cite{Demi,Demi2,Chen} and spin chain systems \cite{Camp1,Takahashi}. Experiments with the counter-diabatic protocol have been demonstrated for control of Bose-Einstein condensates \cite{Bason} and the electron spin of a single nitrogen-vacancy center in diamond \cite{Zhang}. We describe how the addition of a so-called counter-diabatic field (CDF)  to stimulated Raman adiabatic passage (STIRAP) \cite{Gaub} fields improves the efficiency of selective vibrational energy transfer between members of an embedded subset of states of the thiophosgene (SCCl$_2$) molecule and in the HCN$\rightarrow$HNC isomerization reaction. 

In a system with time-dependent Hamiltonian, an adiabatic process is one in which the populations in the instantaneous eigenstates of the Hamiltonian are constant.  That feature is very useful when external field generated variation of the Hamiltonian is used to manipulate the system's evolution.  However, an adiabatic process must be carried out very slowly, at a rate much smaller than the frequencies of transitions between states of the Hamiltonian, and with a driving field intensity that does not induce transitions that compete with the desired population transfer.  Recognition of this restriction has led to the development of control protocols which we call assisted adiabatic transformations; these transformations typically use an auxiliary field to produce, with overall weaker driving fields and/or in a shorter time, and without excitation of competing processes, the desired target state population.  
In its simplest form STIRAP is used to transfer population between states $|1\rangle$  and $|3\rangle$  in a three state manifold in which transitions  $|1\rangle\rightarrow|2\rangle$ and  $|2\rangle\rightarrow|3\rangle$ are allowed but  $|1\rangle\rightarrow|3\rangle$ is forbidden (e.g. see Refs.\cite{Gaub,Coul,Half,Berg,Vita}).  The driving optical field consists of two suitably timed and overlapping laser pulses with the (Stokes) pulse driving the $|2\rangle\rightarrow|3\rangle$  transition preceding the (pump) pulse driving the $|1\rangle\rightarrow|2\rangle$  transition.  The field dressed states of this system are combinations of the bare states  $|1\rangle$ and  $|3\rangle$ with coefficients that depend on the Rabi frequencies of the pump ($\Omega_p$) and Stokes ($\Omega_S$) fields.  Consequently, as those fields vary in time there is an adiabatic transfer of population from $|1\rangle$  to  $|3\rangle$.  Various extended STIRAP methods, involving more than three states, also have been proposed \cite{Kobr1,Kobr2,Kobr3,Kurk2,Kurk1,Torosov}.  In the three state system the efficiency of STIRAP is relatively insensitive to the details of the pulse profile and the pulse separation \cite{Berg}, but complete population transfer requires the product of the pulse overlap $\Delta T$ and pulse Rabi frequencies be sufficiently large,  $\Delta T(\Omega_S^2+\Omega_p^2)^{1/2}>10$, to suppress nonadiabatic transitions that involve  $|2\rangle$ and degrade the efficiency of the population transfer.  Moreover, in real situations, say vibrational energy transfer in a polyatomic molecule, the selected subset of states used for a STIRAP process is embedded in the complete manifold of molecular states.  When the transition dipole moments between the selected states and the other states are not negligible, it is necessary to account for the influence of the background states on the efficiency of the population transfer.  Previous studies of STIRAP generated population transfer in laser-assisted HCN $\rightarrow$ HNC isomerization \cite{Cheng,Jaku2} have revealed that background states coupled to a subset of states used for a STIRAP process degrade the population transfer efficiency.  And, given that the subset of states of interest is embedded in the complete manifold of states, pulsed fields with large amplitude and/or long duration can generate unwanted processes, such as multiphoton ionization, that compete with the wanted population transfer.

In this paper we report studies of state-to-state vibrational energy transfer in an isolated SCCl$_2$ molecule and in the HCN $\rightarrow$ HNC isomerization reaction driven by STIRAP + CDF control.  In the SCCl$_2$ study the efficiency of the controlled and selective population transfer within a subset of states embedded in a dense manifold of states is examined for both the case that the transition dipole moments between the initial and target states are much smaller than the transition dipole moments between background states and the case when they are comparable to the transition dipole moments between background states.  We show that a combination of STIRAP and CDFs that do not individually generate competitive processes can generate greater population transfer efficiency in a subset of states that is coupled to background states than can ordinary STIRAP with field strength and/or pulse duration restricted to avoid excitation of competitive processes.  In the HCN $\rightarrow$ HNC isomerization reaction study we examine the case that the intermediate state of the sequential STIRAP process (see Section IV) has large transition dipole moments to the background states.  We show that the CDF both averts unwanted non-adiabatic population transfer due to failure of the adiabatic condition and enhances the robustness of suppression of the interference with background states.  These calculations are based on the use of an approximation to the CDF, since the exact CDF is too complex to calculate.  We show that the exact CDF for an isolated three level system is a useful approximation to the exact CDF for three and five state systems embedded in background states, and we report a study of the stability of the STIRAP + approximate CDF control protocol to variation of the CDF.

\section{Vibrational Energy Transfer in Thiophosgene}
The SCCl$_2$ molecule has three stretching ($\nu_1, \nu_2, \nu_3$) and three bending ($\nu_4, \nu_5, \nu_6$) vibrational degrees of freedom; the energies and transition dipole moments for all levels in the manifold of vibrational states up to 21,000 cm$^{-1}$ have been reported \cite{Bigw}; see Tables A1 to A3 in the Appendix and \ref{levels_fig}.  The spectrum of states is sufficiently dense, and the distribution of dipole transition moments sufficiently irregular, that complicated dynamics arising from interference with population transfer within a subset of states from the remaining background states is to be expected.  We will focus attention on the efficiency with which population transfer can be selectively directed to one of a pair of nearly degenerate states in the presence of background states.
\begin{figure*}[]
\begin{center}
\includegraphics[width=17.5cm]{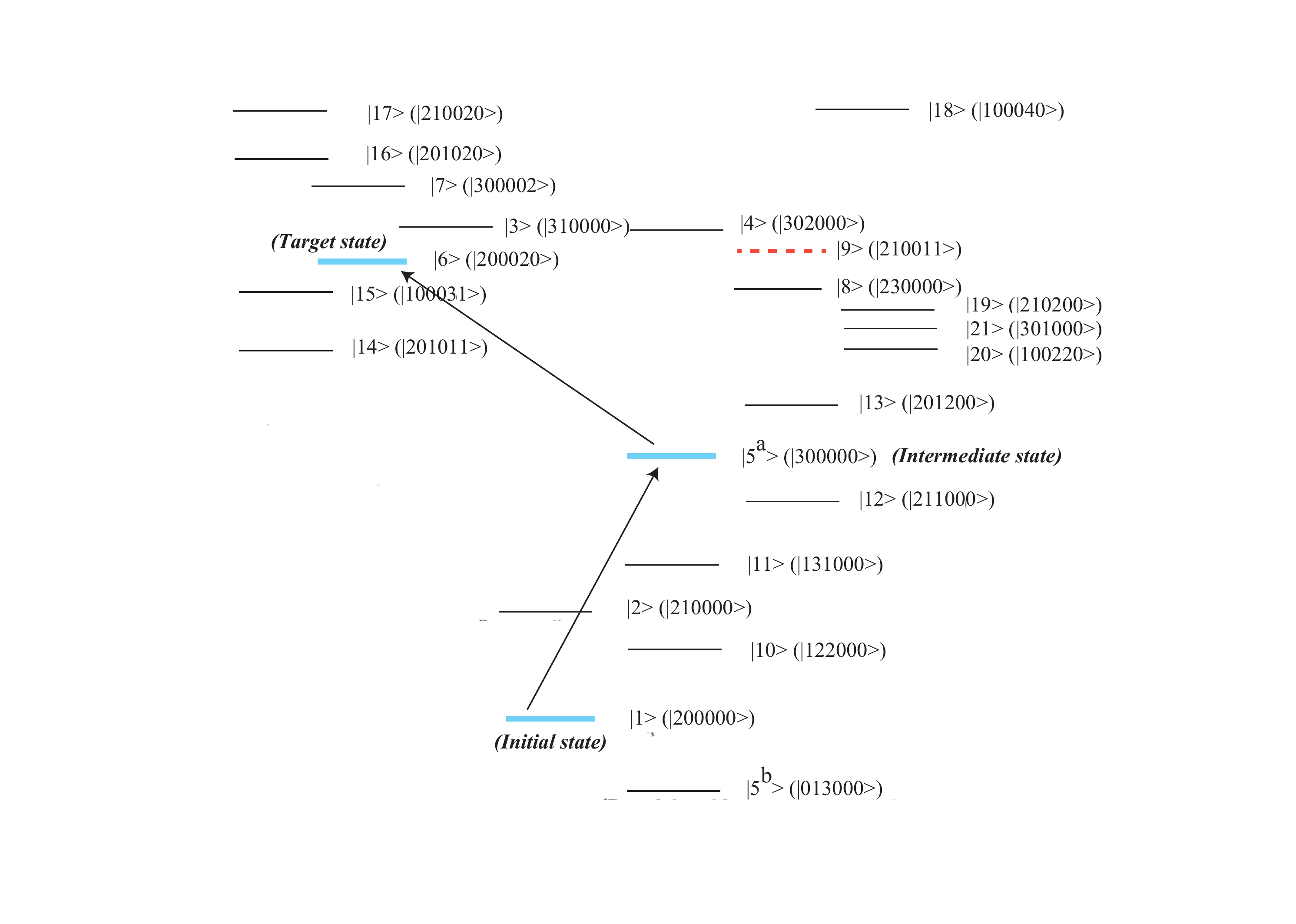}
\end{center}
\caption{
Schematic diagram of the vibrational spectrum of SCCl$_2$.}
\label{levels_fig}
\end{figure*}

First we consider a STIRAP process within a subset of three states ($|200000\rangle, |300000\rangle, |200020\rangle$) decoupled from the full manifold of states.  Hereafter we refer to these three states as  $|1\rangle$, $|5^a\rangle$ and $|6\rangle$, respectively.  The STIRAP process is intended to generate 100$\%$ population transfer from $|200000\rangle$ to $|200020\rangle$.  We note, from \ref{levels_fig} and Table A1, that $|210011\rangle$, hereafter called $|9\rangle$, with energy 5658.1828 cm$^{-1}$, is nearly degenerate with $|6\rangle$, with energy 5651.5617 cm$^{-1}$.  Since the transition moment coupling states $|1\rangle$ and $|6\rangle$ is one order of magnitude smaller than those coupling states $|1\rangle$ and $|9\rangle$ and $|5^a\rangle$ and $|9\rangle$ selective population of $|6\rangle$ poses a problem to be solved.  

Recalling that the STIRAP process transfers population in a system in which the initial and target states are not directly connected by a non-vanishing transition moment, using the interaction representation and the rotating wave approximation (RWA), the Hamiltonian of the three-state system with resonant pump $|1\rangle \rightarrow |5^a\rangle$ and Stokes $|5^a\rangle \rightarrow |6\rangle$ fields can be represented in the form
\begin{eqnarray}
H_{\rm RWA}(t) = -\hbar 
\left( \begin{array}{ccc}
 0 & \Omega_{p}(t) & 0\\
\Omega_{p}(t) & 0 & \Omega_{S}(t)\\
 0 & \Omega_{S}(t) & 0
\end{array}\right),
\label{HRWA}
\end{eqnarray}
with $\Omega_p$ and $\Omega_S$ the Rabi frequencies defined by
\begin{equation}
\begin{minipage}[c]{0.80\linewidth}
\begin{eqnarray}
\Omega_{p}(t) &=& \frac{\mu_{15^a}E_{p}^{(e)}(t)}{2\hbar},\nonumber\\
\Omega_{S}(t) &=& \frac{\mu_{5^a6}E_{S}^{(e)}(t)}{2\hbar},\nonumber
\end{eqnarray}
\end{minipage}
\label{eq_omegas}
\end{equation}
%
where $E_{p(S)}^{(e)}(t)$ is the envelope of the amplitude of the pump (Stokes) 
field and $\mu_{ij}$ the transition dipole moment between states
$|i\rangle$ and $|j\rangle$.
Note that, by assumption, $\mu_{16}=0$.  The time-dependent field-dressed eigenstates of this system are linear combinations of the field-free states with coefficients that depend on the Stokes and pump field magnitudes and the transition dipole moments. The field-dressed state of interest to us is
\begin{eqnarray}
|a_0(t)\rangle = \cos\Theta(t) |1\rangle - \sin\Theta(t) |6\rangle,
\end{eqnarray}
where
\begin{equation}
\tan\Theta(t) = \frac{\Omega_{p}(t)}{\Omega_{S}(t)}.
\label{theta1}
\end{equation}
Because the Stokes pulse precedes but overlaps the pump pulse, initially $\Omega_p \ll \Omega_S$ and all of the population is initially in field-free state $|1\rangle$.  At the final time $\Omega_p \gg \Omega_S$ so all of the population in $|a_0(t)\rangle$ projects onto the target state $|6\rangle$.  Note that $|a_0(t)\rangle$ has no projection on the intermediate field-free state $|5^a\rangle$.  Suppose now that either the pulsed field duration or the field strength must be restricted to avoid exciting unwanted process that compete with the desired population transfer, with the consequence that the condition $\Delta T (\Omega_S^2 + \Omega_p^2)^{1/2} > 10$ cannot be met.  This situation prompts us to seek a CDF that restores the adiabatic population transfer.  The counter-diabatic Hamiltonian for the STIRAP process can be represented as \cite{Demi}
\begin{eqnarray}
H_{\rm CD}(t) = \hbar 
\left( \begin{array}{ccc}
 0 & 0 & i\dot{\Theta}\\
 0 & 0 & 0\\
 -i\dot{\Theta} & 0 & 0
\end{array}\right).
\end{eqnarray}
By inversion of the RWA the counter-diabatic field is found to be
\begin{equation}
E_{\rm CD}(t) = \mbox{sgn}(\ep_6-\ep_1) \frac{2\hbar\dot{\Theta}(t)}{\mu_{16}}
\sin (\omega_{\rm CD}t),
\label{ECD1}
\end{equation}
\begin{equation*}
\omega_{\rm CD} = \frac{|\ep_6-\ep_1|}{\hbar},
\end{equation*}
where $\ep_i$ is the energy of $|i\rangle$.  This CDF, when added to the original STIRAP fields, will generate the desired $|1\rangle\rightarrow |6\rangle$ adiabatic population transfer in the three-state system.  However, as noted earlier, when there are background states coupled to the three-state subsystem, strong counter-diabatic and STIRAP fields can also generate processes that compete with population transfer.  We now show that the combined STIRAP and CDFs that do not individually generate competitive processes can generate greater $|1\rangle\rightarrow |6\rangle$ population transfer efficiency in a subset of states that is coupled to background states than can ordinary STIRAP with field strength and/or pulse duration that are restricted to avoid excitation of competitive processes.  Because the exact CDF for the full system involving the subset states embedded in the background states is very complex we use \ref{ECD1} as an approximation to the exact CDF for the system being discussed.  Then it is also useful to examine the stability of the driven population transfer to variation of the approximate CDF.

\section{Numerical Results}
We choose the envelopes of the amplitude of the pump and Stokes fields to be
\begin{eqnarray}
E_{p(S)}^{(e)}(t) = \tilde{E}_{p(S)} 
\exp\Big{[}-\frac{(t-T_{p(S)})^2}{(\Delta \tau)^2}
\Big{]},
\label{eq_laser1}
\end{eqnarray}
where $\Delta \tau = \mbox{FWHM} / (2\sqrt{\ln 2}) $, and 
FWHM is the full width at half maximum of the Gaussian pulse with  
maximum intensity $\tilde{E}_{p(S)}$ that is centered at $T_{p(S)}$.  
Using \ref{eq_omegas,theta1} $\dot\Theta$  takes the form
\begin{eqnarray}
&&\dot\Theta = 
\frac{\Omega_{S}\dot\Omega_{p}-\dot\Omega_{S}\Omega_{p}}{
\Omega_{S}^2 + \Omega_{p}^2} \nonumber\\
&&= 
\frac{2\tilde E_{p}\tilde E_{S}\mu_{p}\mu_{S}}{(\Delta\tau)^2}(T_{p}-T_{S})\nonumber\\
&&\times\frac{\exp\Big{[}-\frac{1}{(\Delta\tau)^2}\Big\{(t-T_{p})^2
  +(t-T_{S})^2\Big\}\Big{]}}{\tilde E_{p}^2\mu_{p}^2\exp\Big{[}
-\frac{2(t-T_{p})^2}{(\Delta\tau)^2}\Big{]} + 
\tilde E_{S}^2\mu_{S}^2\exp\Big{[}
-\frac{2(t-T_{S})^2}{(\Delta\tau)^2}\Big{]}
}.\nonumber
\label{thetadot}
\end{eqnarray}
\begin{equation}
\label{dTheta}
\end{equation}
From \ref{ECD1} we see that the intensity of the CDF is proportional to $\dot\Theta$.  For given $\tilde{E}_{p(S)}$ the peak intensity of the CDF is proportional to $(T_p-T_S)/\mbox{FWHM}^2$, because the final factor in \ref{dTheta} does not affect the peak value of $\dot\Theta$.  In the numerical calculations we solved the time-dependent Schr$\ddot{\mbox{o}}$dinger equation with a fourth order Runge-Kutta integrator in a basis of bare matter eigenstates with $T_p-T_S = \mbox{FWHM} / (2\sqrt{\ln 2})$ and the intensities and widths shown in \ref{table_laser1}.
\begin{table}[htb]
  \caption{Pump and Stokes laser parameters}
  \begin{tabular}{|l|c|c|}
    \hline
         & $\tilde{E}_{p(S)}$ (a.u.)  & FWHM (ps) \\
    \hline
     Pump pulse    & $3.11\times 10^{-6}$ & 215 \\
     Stokes pulse  & $3.44\times 10^{-6}$ & 215 \\
\hline
  \end{tabular}
\label{table_laser1}
\end{table}
The transition dipole moments used in the calculations are shown in Tables A2 and A3. All transitions between all of the states shown were included in the calculations.  The transition dipole moments relevant for these calculations are $\mu_{15^a} = 0.2062$, $\mu_{5^a6} = 0.2090$, $\mu_{16}= 0.0345$ and $\mu_{5^a9}= 0.4138$.  We note again that the transition dipole moment between the initial and the target states, $\mu_{16}$, is one order of magnitude smaller than $\mu_{15^a}$, $\mu_{5^a6}$, and $\mu_{5^a9}$.

\begin{figure}[h!]
\begin{center}
\includegraphics[width=8cm]{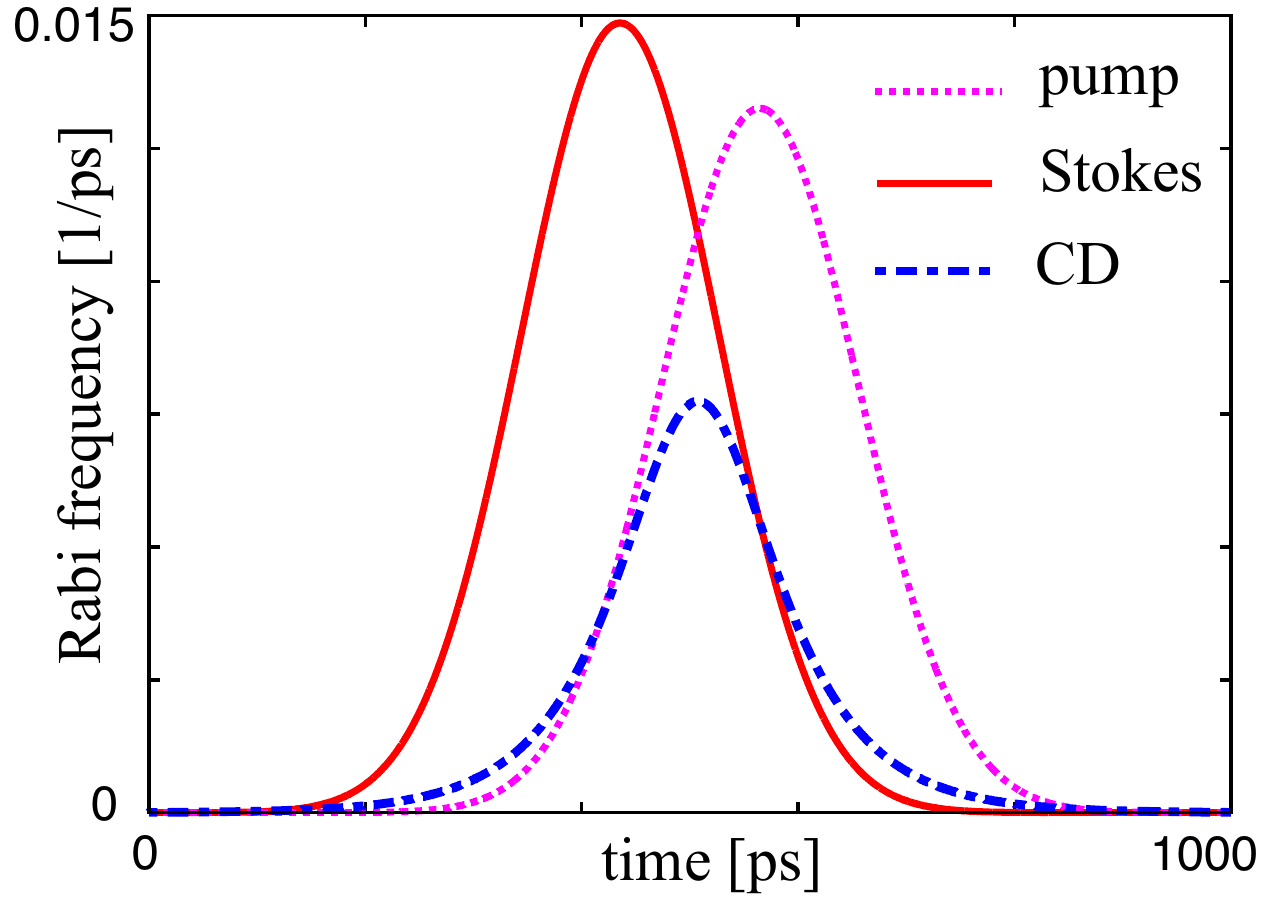}
\end{center}
\caption{
Time-dependence of the Rabi frequencies of the pump and Stokes and 
CDFs.}
\label{Omega_215}
\end{figure}

\begin{figure}[h!]
\begin{center}
\includegraphics[width=8cm]{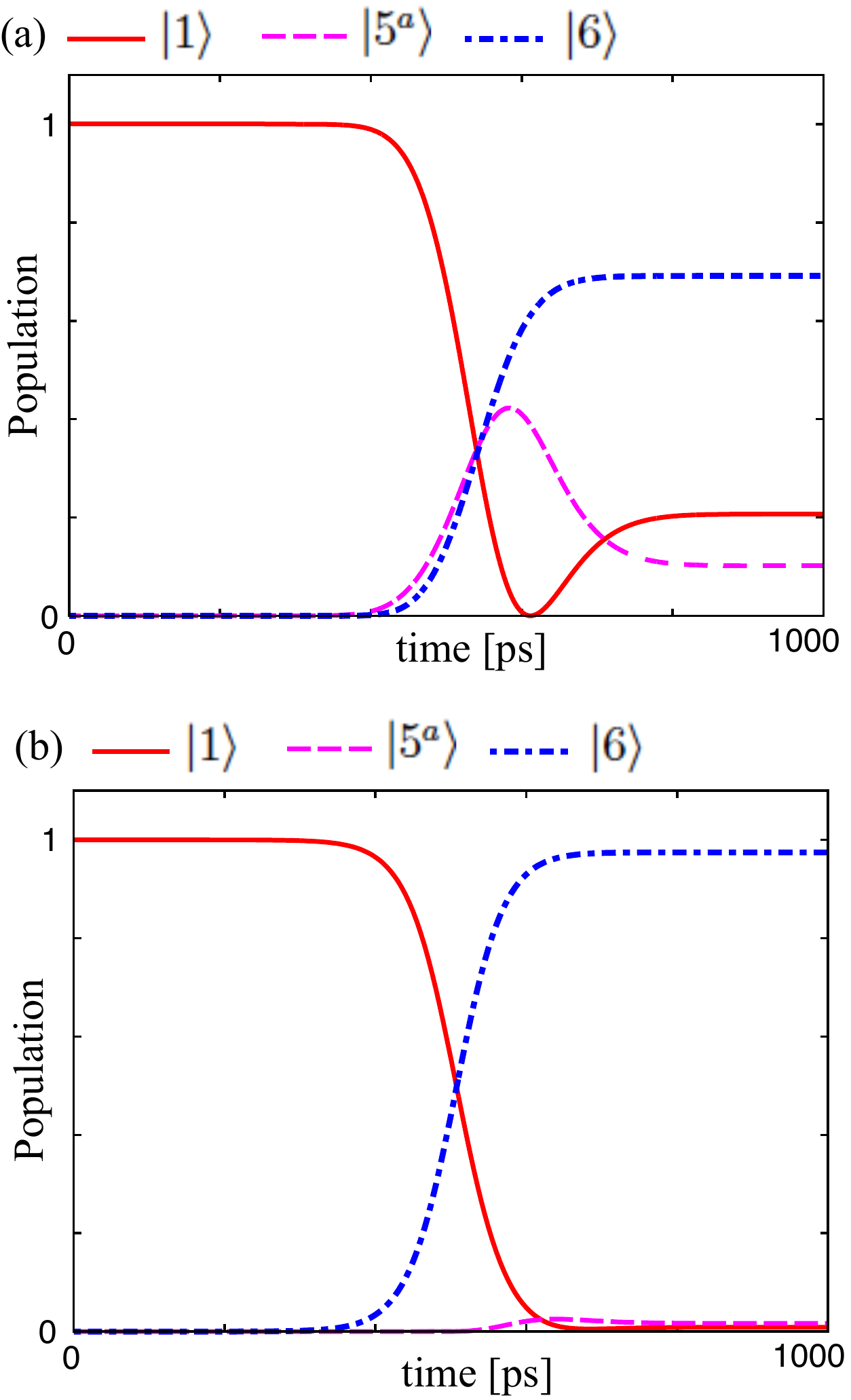}
\end{center}
\caption{
Time-dependence of the population of the initial, 
intermediate and the target states for (a) STIRAP ($\lambda=0$) and 
(b) STIRAP + CDF control for $\lambda=1$.}
\label{p_stirap_com_215}
\end{figure}

So as to study the stability of the efficiency of the population transfer driven 
by a variable CDF we represent the total driving field in the form
\begin{equation}
E(t) = E_p(t) + E_S(t) + \lambda E_{\rm CD}(t),
\label{E_lambda}
\end{equation}
where $\lambda=0$ corresponds to driving the system with only the 
STIRAP fields and $\lambda=1$ to driving the system with the STIRAP and the CDFs.  
The time-dependences of the Rabi frequencies $\Omega_p(t)$, $\Omega_S(t)$
and $\Omega_{\rm CD}(t) = d\Theta/dt$ are shown 
in \ref{Omega_215}. 
The peak of the CDF time dependent Rabi frequency is located between those of the Stokes and the pump fields, and the CDF pulse overlaps both.  The time-dependence of the population of the initial, intermediate and target states for (a) pure STIRAP ($\lambda=0$) and (b) STIRAP + CDF control ($\lambda=1$) are exhibited in \ref{p_stirap_com_215}.  The populations of the other states are negligible.  The fidelity of the population transfer, defined as the population of the target state at the final time, is 0.974 for STIRAP + CDF control, but only 0.688 for STIRAP control because of the failure of the adiabatic condition and the interference with the background states. 

\begin{figure}[h!]
\begin{center}
\includegraphics[width=8cm]{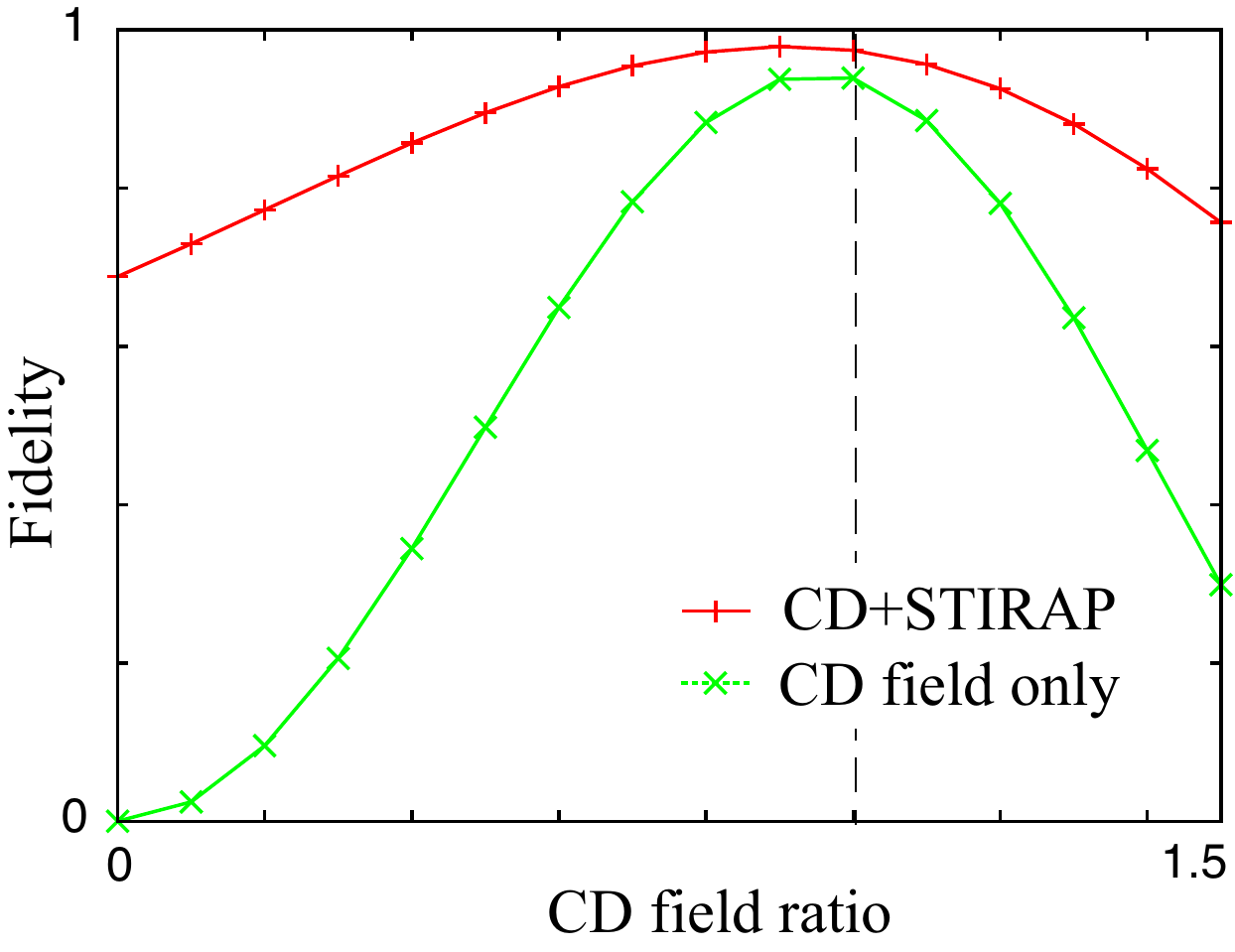}
\end{center}
\caption{
Comparison of the fidelity of STIRAP + CDF control and CDF alone control with FWHM$=215$ ps and $T_{p}-T_{S} = \mbox{FWHM} / (2\sqrt{\ln 2})$ for various
 $\lambda$  in the population transfer $|1\rangle
\rightarrow |6\rangle$.}
\label{fid_cd_com}
\end{figure}

\begin{figure}[h!]
\begin{center}
\includegraphics[width=8cm]{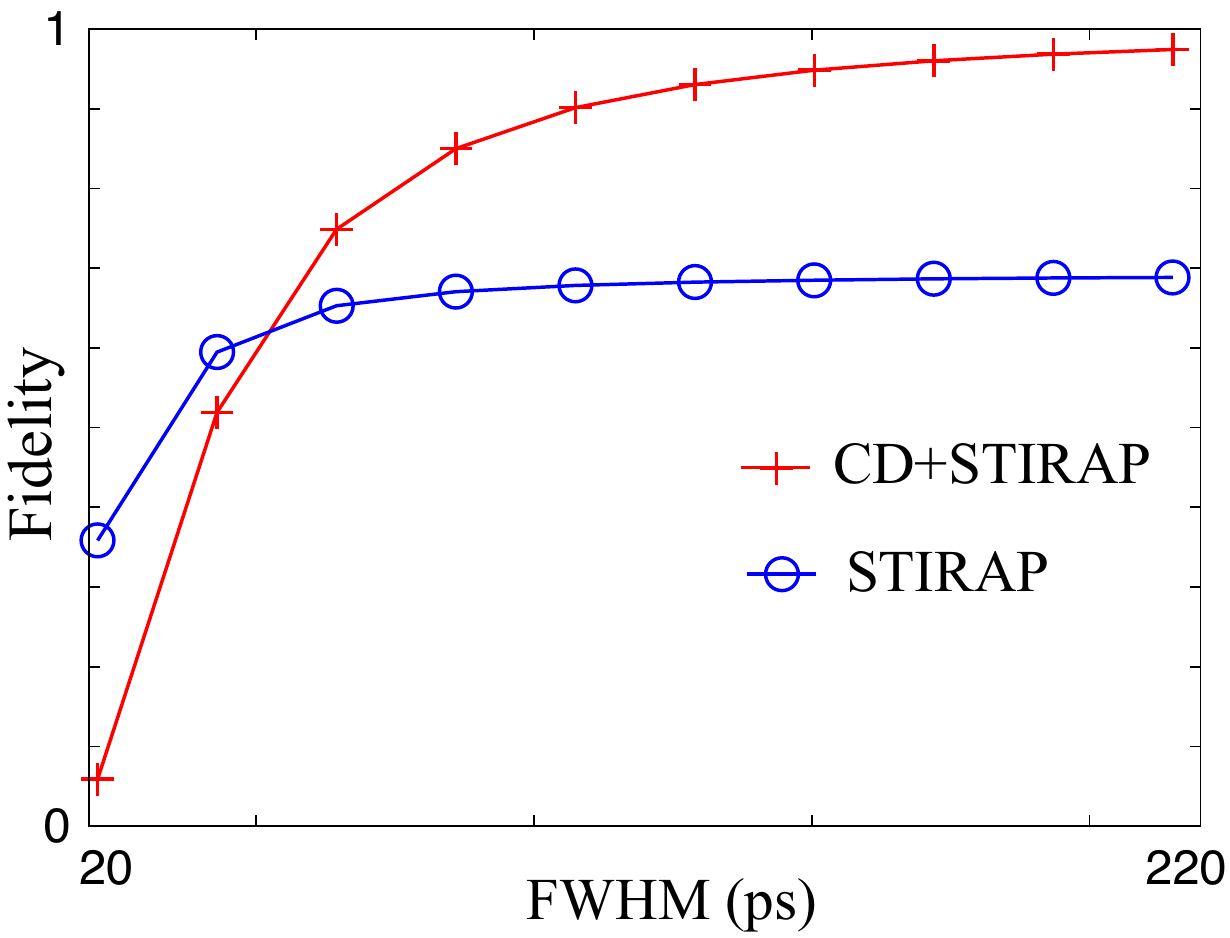}
\end{center}
\caption{
Comparison of the FWHM-dependence of fidelity between STIRAP + CDF control 
and STIRAP with $T_{p}-T_{S} = \mbox{FWHM} / (2\sqrt{\ln 2})$ in the population transfer $|1\rangle
\rightarrow |6\rangle$.
}
\label{fid_FWHM}
\end{figure}

Because the CDF pulse has an area equal to $\pi$ \cite{Chen}, in a two-level system it alone can generate perfect population transfer to the target state.  However the efficiency of such simple single pulse control is degraded due to the interaction with the background states, and is not stable to variation of the area of the pulse. The stability of STIRAP + CDF control and CDF alone control to variation of the amplitude of the CDF, characterized by the value of $\lambda$, is monitored by using the fidelity.  The $\lambda$-dependence of the fidelity is shown in \ref{fid_cd_com} for STIRAP + CDF control and CDF 
control.  The decreased sensitivity to variation of $\lambda$ of STIRAP + CDF control relative to CDF control is evident.  Note that even when the applied CDF is weaker than the $\lambda=1$ CDF, STIRAP + CDF control generates higher fidelity than either does individually. The results in \ref{fid_cd_com} show that the fields associated with STIRAP and CDF generated population transfer are complementary.  The value of $\lambda$ that corresponds to the peak of the fidelity is slightly smaller than 1 in \ref{fid_cd_com}, because of interference with the background states generated by the strong CDF.  \ref{dTheta} and $T_p-T_S = \mbox{FWHM} / (2\sqrt{\ln 2})$ require that the CDF strength increases with decreasing FWHM, with the peak intensity proportional to FWHM$^{-1}$.  Thus for short pulses the CDF intensity can be great enough to generate unwanted excitations.  
 \ref{fid_FWHM} shows the dependence of the fidelity on the FWHM for STIRAP + CDF and STIRAP control of population transfer. For each FWHM, the values of $\tilde{E}_{p(S)}$ are adjusted so that the pulse areas of the pump and Stokes fields are the same as those in  \ref{Omega_215}.  The fidelity of STIRAP is insensitive to the pulse FWHM for sufficiently large FWHM because of the constant area of the pump and Stokes pulses and the small field amplitudes.  However for small pulse FWHM a decrease of the fidelity is seen as the FWHM decreases because the strong pump and Stokes fields enhance the interference with the background states.  The fidelity of the STIRAP + CDF control is higher than that of STIRAP for pulses with FWHM $>$ 50 ps.  The drop of the fidelity in the STIRAP + CDF control is more rapid than STIRAP because the CDF is much stronger than the pump and Stokes fields due to the small magnitude of $\mu_{16}$.

\begin{figure}[h!]
\begin{center}
\includegraphics[width=8cm]{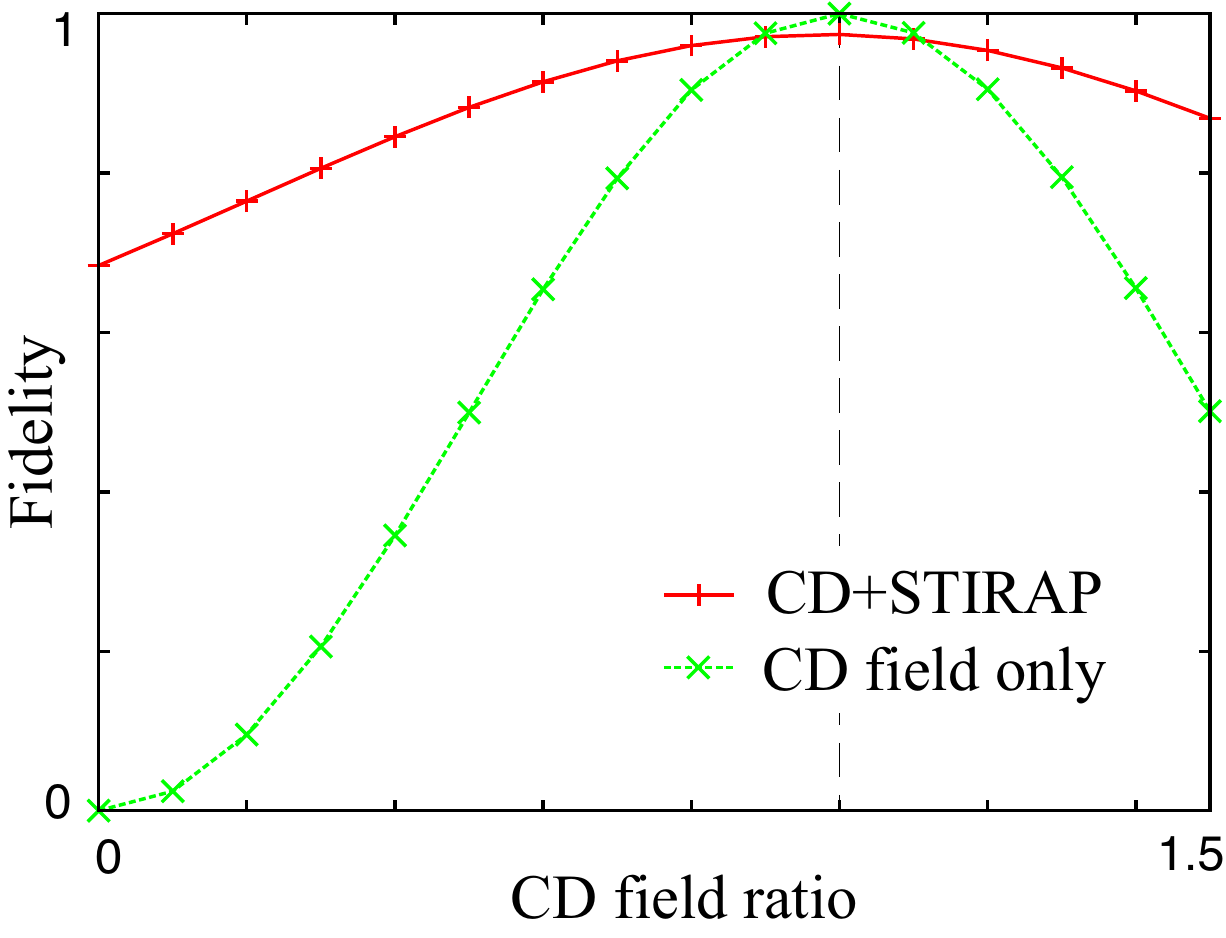}
\end{center}
\caption{
Comparison of the fidelity of STIRAP + CDF control and CDF alone control  with FWHM$=215$ ps and $T_{p}-T_{S} = \mbox{FWHM} / (2\sqrt{\ln 2})$ for various $\lambda$ in the population transfer $|1\rangle
\rightarrow |3\rangle$.
}
\label{fid_cd_dep_123}
\end{figure}

\begin{figure}[h!]
\begin{center}
\includegraphics[width=8cm]{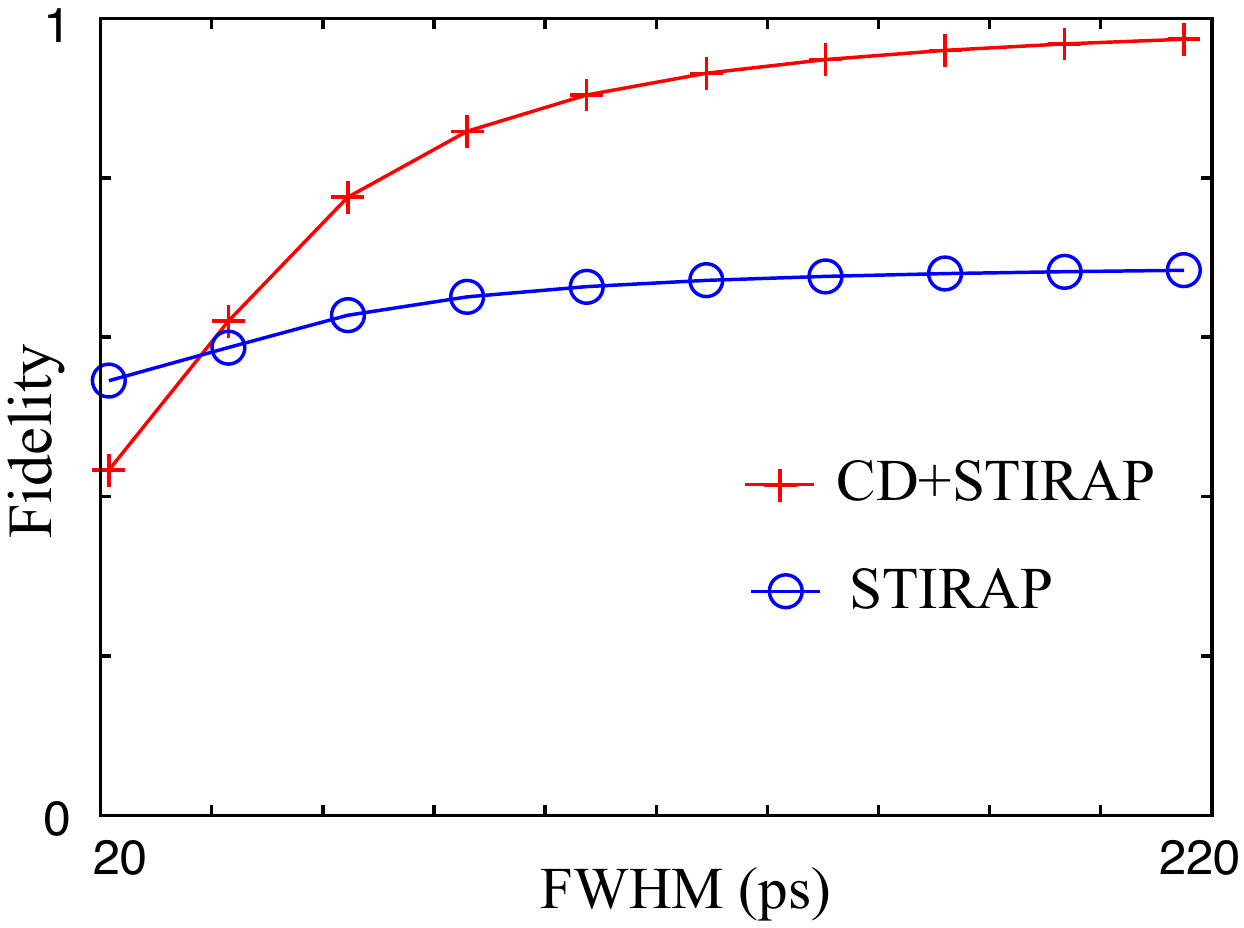}
\end{center}
\caption{ 
Comparison of the FWHM-dependence of fidelity between STIRAP + CDF control 
and STIRAP with $T_{p}-T_{S} = \mbox{FWHM} / (2\sqrt{\ln 2})$ in the population transfer $|1\rangle
\rightarrow |3\rangle$.
}
\label{fid_com_1_2_3}
\end{figure}

 Next we take a different set of states, namely  $|1\rangle$, $|2\rangle$ and $|3\rangle$, as the initial, intermediate and target states of a STIRAP process.  The background states,  $|4,6,7,9\rangle$, have the energy close to the target state energy.  In contrast to the previous example, the transition dipole moment between the initial and target states is larger than those between the initial and intermediate states and the intermediate and target states; the CDF is smaller than the pump and Stokes fields due to the large transition dipole moment $\mu_{13}$.  We choose the amplitude of the pump and Stokes fields so that the corresponding Rabi frequencies are the same as in \ref{fid_cd_com,fid_FWHM} for each $\lambda$ or FWHM.  The $\lambda$-dependence of the fidelity is shown in \ref{fid_cd_dep_123} for STIRAP + CDF control and CDF control. 
 The CDF alone can generate high fidelity, indeed almost unity, if the pulse area is $\pi$, because the $\lambda=1$ CDF is not strong.  However, the efficiency of this population transfer is not stable to variation of the field strength.  On the other hand, STIRAP + CDF control generates higher fidelity than does CDF control for a wide range of $\lambda$, revealing the complementarity of the STIRAP and CDF fields.  \ref{fid_com_1_2_3} shows the dependence of the fidelity on the FWHM in the STIRAP + CDF and STIRAP control processes.  For each FWHM, $\tilde{E}_{p(S)}$ are adjusted so that the pulse areas corresponding to the pump and Stokes fields are the same as in \ref{Omega_215}. 
 The decrease of the fidelity of the STIRAP + CDF control process with decreasing FWHM is rapid compared that of STIRAP because of the addition of the CDF. However it is not rapid compared to that shown in \ref{fid_FWHM} because the CDF is smaller than that used in the calculations displayed in \ref{fid_FWHM}.  Clearly, complementarity of the CDF and STIRAP fields is retained if unwanted excitations do not occur.

\section{HCN $\rightarrow$ HNC Isomerization Reaction}
We now study the HCN $\rightarrow$ HNC isomerization process driven by STIRAP + CDFs. We consider the case that some of the background states are strongly coupled to the intermediate state, so that the coupling degrades the efficiency of the STIRAP control even if the adiabatic condition is met.  Although the CDF is designed to avert unwanted non-adiabatic population transfer, the CDF directly couples the initial and target states and thereby also makes the control less sensitive to the influence of background states than is the ordinary STIRAP control.

\begin{figure}[h!]
\begin{center}
\includegraphics[width=7.5cm]{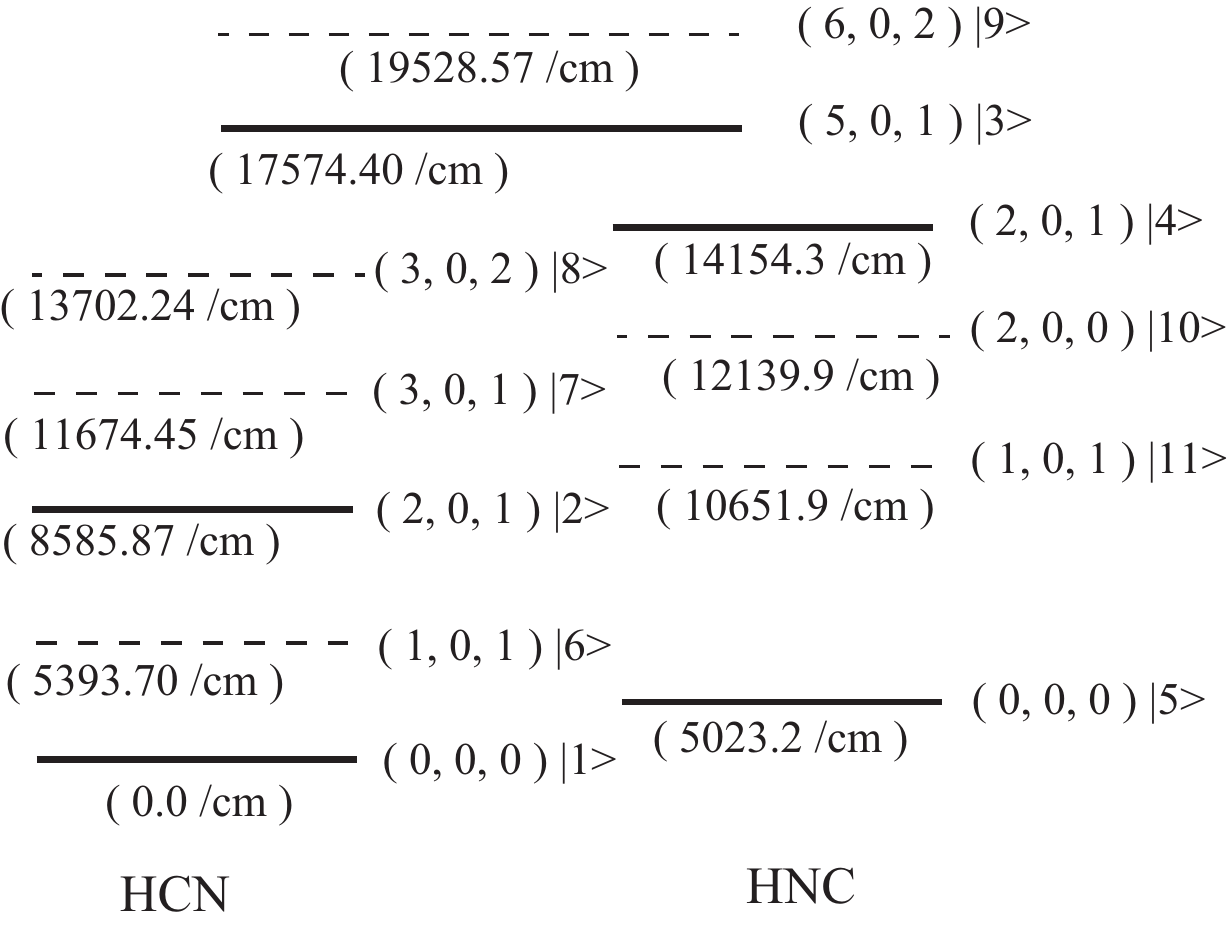}
\end{center}
\caption{
Schematic diagram of the vibrational spectrum of states used for the numerical simulations.  The states selected for use in the successive STIRAP processes are represented with thick lines, and the background states are represented with thin dashed lines.}
\label{fig_lavels}
\end{figure}

The three-dimensional potential energy surface for non-rotating HCN/HNC has been well studied \cite{Smith,Yang,Jonas,Bowm1,Jaku}.  The key degrees of freedom that characterize this surface are the CH, NH and CN stretching motions and the HNC bending motion. These are combined in the symmetric stretching, bending and asymmetric stretching normal modes, with quantum numbers ($\nu_1$,$\nu_2$,$\nu_3$), respectively.  The vibrational energy levels of HCN and HNC have been calculated by Bowman et al \cite{Bowm1}.  Driving the HCN $\rightarrow$ HNC isomerization with a simple monochromatic laser field is difficult because the Franck-Condon factors between the ground vibrational states (0,0,0) of HCN and HNC and the vibrational levels close to the top of the isomerization barrier (5,0,1) are extremely small.  In the model system considered by Kurkal and Rice \cite{Kurk1} eleven vibrational states, shown schematically in \ref{fig_lavels}, are considered; rotation of the molecule is neglected.  They proposed overcoming the Franck-Condon barrier with sequential STIRAP, consisting of two successive STIRAP processes.  The use of this sequence is intended to avert the unwanted competition with other processes that can be generated by the very strong fields that would be needed to overcome the Franck-Condon barriers encountered in a single STIRAP process.  Of course, to be realistic, the fields used in the sequential STIRAP driven isomerization also must be small enough to not generate unwanted competing processes.  In the first step of the sequential STIRAP, the (0,0,0), (2,0,1) and (5,0,1) states of HCN are chosen as the initial, intermediate and final states, respectively; in the second STIRAP process, the (5,0,1), (2,0,1) and (0,0,0) states of HNC are taken as the initial, intermediate and final states, respectively.  Other states, shown with dashed lines in \ref{fig_lavels}, are regarded as background states. 
The pump 1 field is resonant with the transition from the (0,0,0) state of HCN to the (2,0,1) state of HCN; the Stokes 1 field is resonant with the transition from the (2,0,1) state of HCN to the (5,0,1) state; the pump 2 field is resonant with the transition from the (5,0,1) state to the (2,0,1) state of HNC; and the Stokes 2 field is resonant with the transition from the (2,0,1) state to the (0,0,0) state of HNC.  The transition dipole moments and the energy of the vibrational states denoted by $|i\rangle$ are listed in Tables A4 and A5 in the Appendix.

\begin{figure}[h!]
\begin{center}
\includegraphics[width=8cm]{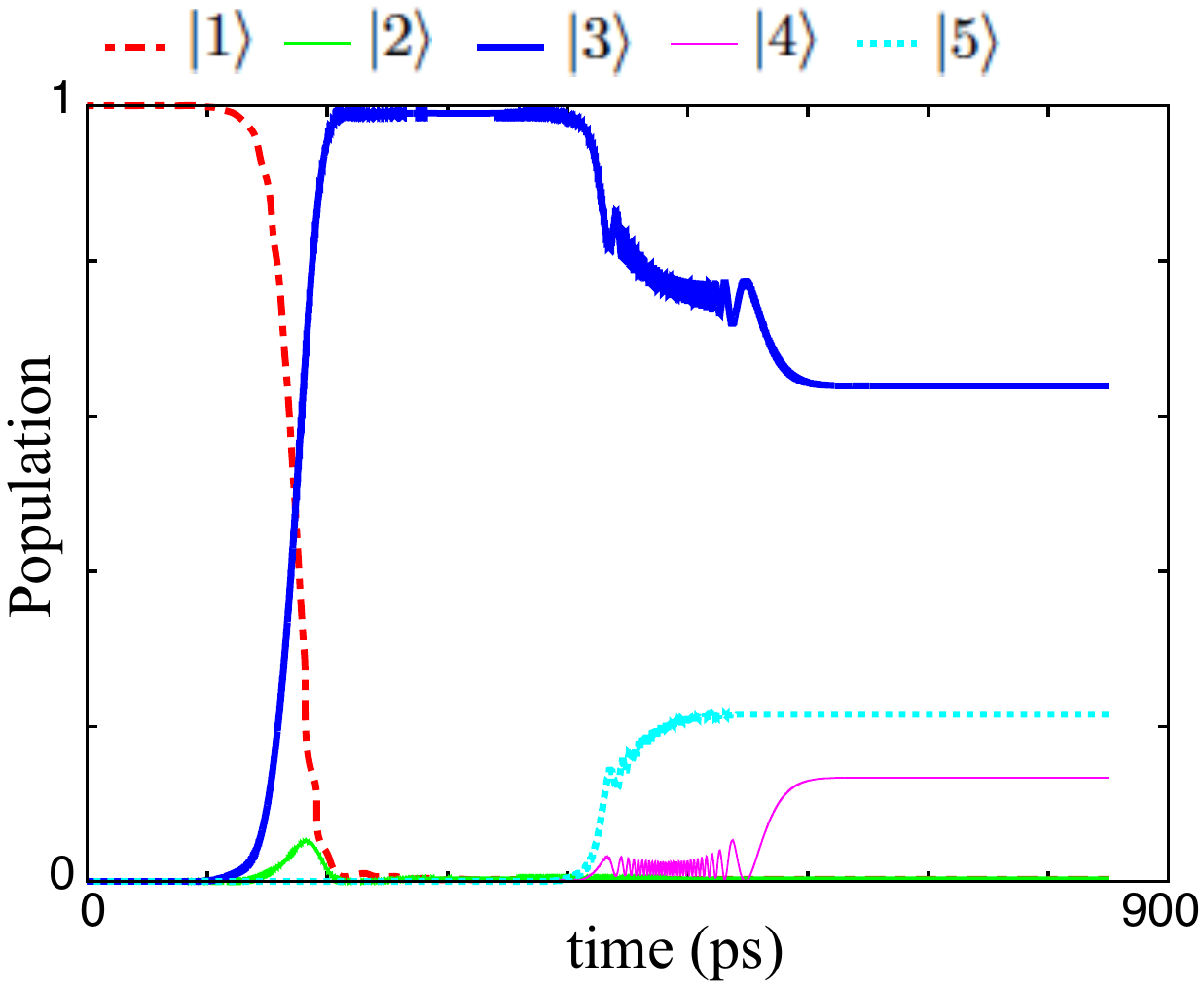}
\end{center}
\caption{Time-dependence of the population for $|1-5\rangle$ in the sequential STIRAP.}
\label{sequential_STIRAP}
\end{figure}

Kurkal and Rice showed that the first STIRAP process is not sensitive to coupling with the background states caused by the Stokes 1 pulse \cite{Kurk1}.  However the second STIRAP process is influenced by interference with the background states because the intermediate state of the second STIRAP process has large transition dipole moments between the background states.  The time-dependences of the populations of states $|1\rangle-|5\rangle$ in the sequential STIRAP process are displayed in \ref{sequential_STIRAP}.  We take the field strengths of the pump and the Stokes fields to be
\begin{eqnarray}
E_{j,p(S)}^{(e)} = \tilde{E}_{j,p(S)} \exp\Big{[}-\frac{(t-T_{j,p(S)})^2}{(\Delta \tau)^2}\Big{]}
\label{eq_laser1}
\end{eqnarray}
where  $\Delta \tau = \mbox{FWHM} / (2\sqrt{\ln 2}) $, and FWHM is the full width at half maximum of the Gaussian pulse with maximum intensity $\tilde{E}_{j,p(S)}$ that is centered at  $T_{j,p(S)}$ and $j=(1,2)$ denotes the first ($j = 1$) and the second STIRAP ($j = 2$) process.  The parameters of the laser fields used in our calculations are shown in \ref{table_laser_seq}.  It is seen clearly in \ref{sequential_STIRAP} that the first STIRAP in the sequential STIRAP process is robust with respect to interference from the background states.  For that reason we assume that the population is transferred from $|1\rangle$ to $|3\rangle$ completely, and we focus attention on the second STIRAP process, choosing $|3\rangle$ as the initial state of the STIRAP + CDF control process.  States $|1\rangle$ and $|2\rangle$ are now and hereafter regarded as background states.
\begin{table}[htb]
  \caption{Pump 1, 2 and Stokes 1, 2 laser parameters}
  \begin{tabular}{|l|c|c|c|}
    \hline
         & $\tilde{E}_{j,p(S)}$ (a.u.)  & $T_{j,p(S)}$ & FWHM (ps) \\
    \hline
     Stokes 1   & 0.00692 & 133 & 85 \\
     pump 1 & 0.00728 & 194 & 85 \\
     Stokes 2   & 0.00575 & 423 & 85 \\
     pump 2 & 0.00220 & 484 & 85 \\
\hline
  \end{tabular}
\label{table_laser_seq}
\end{table}

\section{Numerical Results}
The parameters descriptive of the pump 2 and Stokes 2 fields used in the STIRAP + CDF control process are shown in \ref{table_laser}. $\tilde{E}_{2,p}$ and $\tilde{E}_{2,S}$ were chosen so that the peaks of the Rabi frequencies of the pump and the Stokes fields are the same. The intensity of the field is defined by \cite{Berg}
\begin{eqnarray}
I(t) = \frac{1}{2}|E^{(e)}(t)|^2 c  \ep_0,
\end{eqnarray}
with peak values 3.03$\times 10^{10}$ W/cm$^2$ and 2.90 $\times 10^{11}$ W/cm$^2$ for the pump 2 and Stokes 2 fields, respectively.  
\begin{table}[htb]
  \caption{Pump 2 and Stokes 2 laser parameters}
  \begin{tabular}{|l|c|c|}
    \hline
         & $\tilde{E}_{2,p(S)}$ (a.u.)  & FWHM (ps) \\
    \hline
     pump 2   & 0.0009295 & 212.5 \\
     Stokes 2 & 0.002875 & 212.5 \\
\hline
  \end{tabular}
\label{table_laser}
\end{table}
The CDF is represented as
\begin{eqnarray}
E_{\rm CD}(t) &=& \mbox{sgn}(\ep_5-\ep_3) \frac{2\hbar\dot{\Theta}(t)}{\mu_{35}}
\sin (\omega_{\rm CD}t),\nonumber\\
\omega_{\rm CD} &=& \frac{|\ep_3-\ep_5|}{\hbar}.
\label{ECD2}
\end{eqnarray}

\begin{figure}[h!]
\begin{center}
\includegraphics[width=8cm]{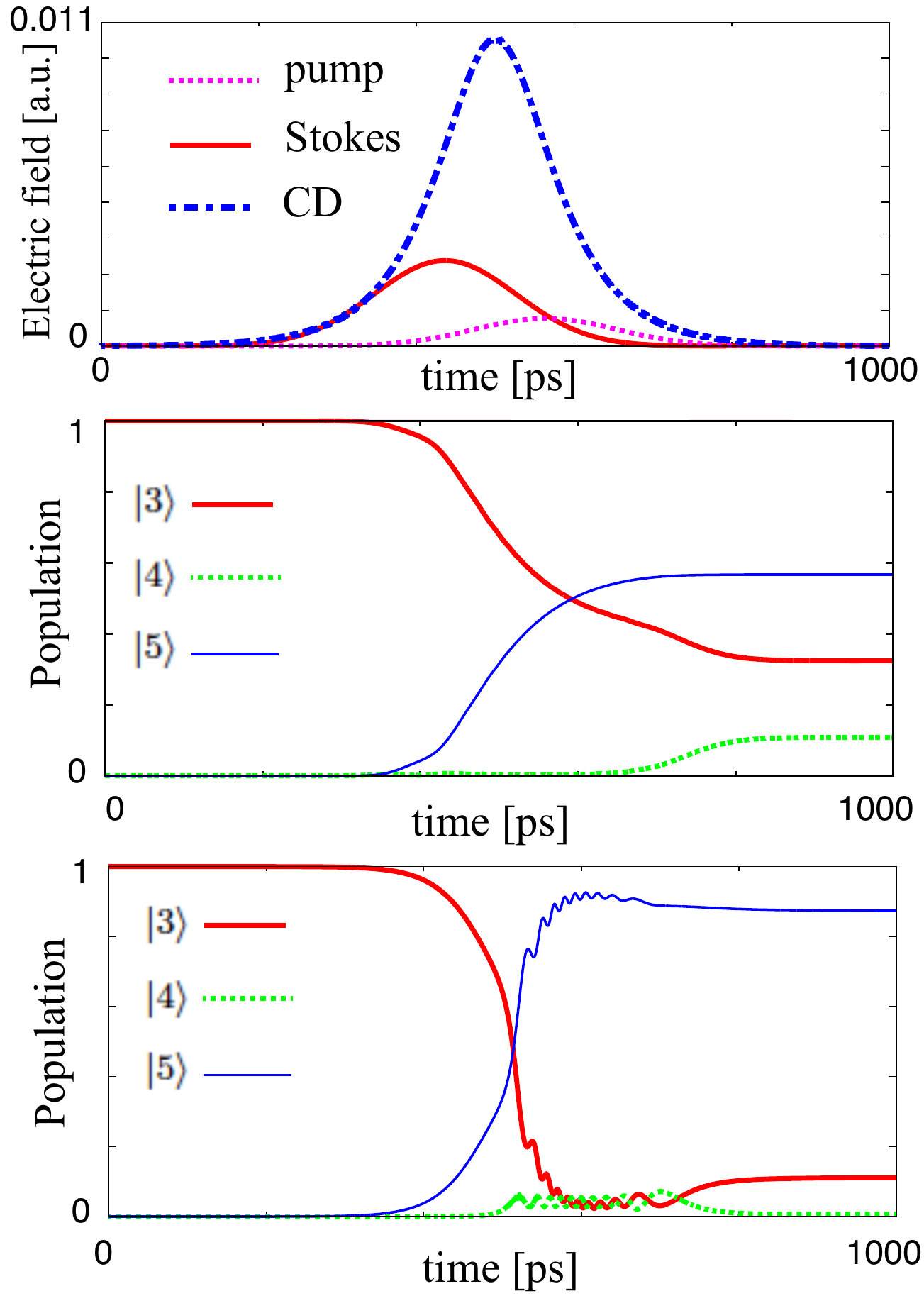}
\end{center}
\caption{
Time-dependence of the STIRAP and CDFs for FWHM$=$212.5 ps
(top panel). Time-dependences of the populations of 
$|3\rangle, |4\rangle, |5\rangle$ for STIRAP (middle panel) 
and STIRAP + CDFs (bottom panel)
with FWHM$=$212.5 ps and $T_{2p}-T_{2S} = \mbox{FWHM} / (2\sqrt{\ln 2})$}.
\label{E_p_FWHM5}
\end{figure}

The time-dependences of the pump, the Stokes and the CDFs for FWHM = 212.5 ps are shown in the top panel of \ref{E_p_FWHM5} for the case that $T_{2p}-T_{2S} = \mbox{FWHM} / (2\sqrt{\ln 2})$.  The time-dependences of the populations of $|3\rangle$, $|4\rangle$ and $|5\rangle$ in the STIRAP and STIRAP + CDF control process for the case that FWHM = 212.5 ps are shown in the middle and bottom panels in \ref{E_p_FWHM5}, respectively. 
The populations of the background states are negligible.  Note that the STIRAP process with FWHM = 212.5 ps generates an isomerization yield with fidelity (final population for $|5\rangle$) about 0.995 if the three states $|3\rangle$, $|4\rangle$ and $|5\rangle$ are decoupled from the background states.  However when $|3\rangle$, $|4\rangle$ and $|5\rangle$ are coupled to the background states the fidelity decreases to about 0.57.  On the other hand, when the CDF is coupled with the pump and Stokes fields the fidelity recovers to about 0.88. We emphasize that because the CDF couples the initial and target states it helps the direct population transfer from the initial state to the target state and that STIRAP + CDF control is less sensitive than is the ordinary STIRAP to interference with the background states coupled to the intermediate state if the amplitude of the field is not too large.  

\begin{figure}[h!]
\begin{center}
\includegraphics[width=8cm]{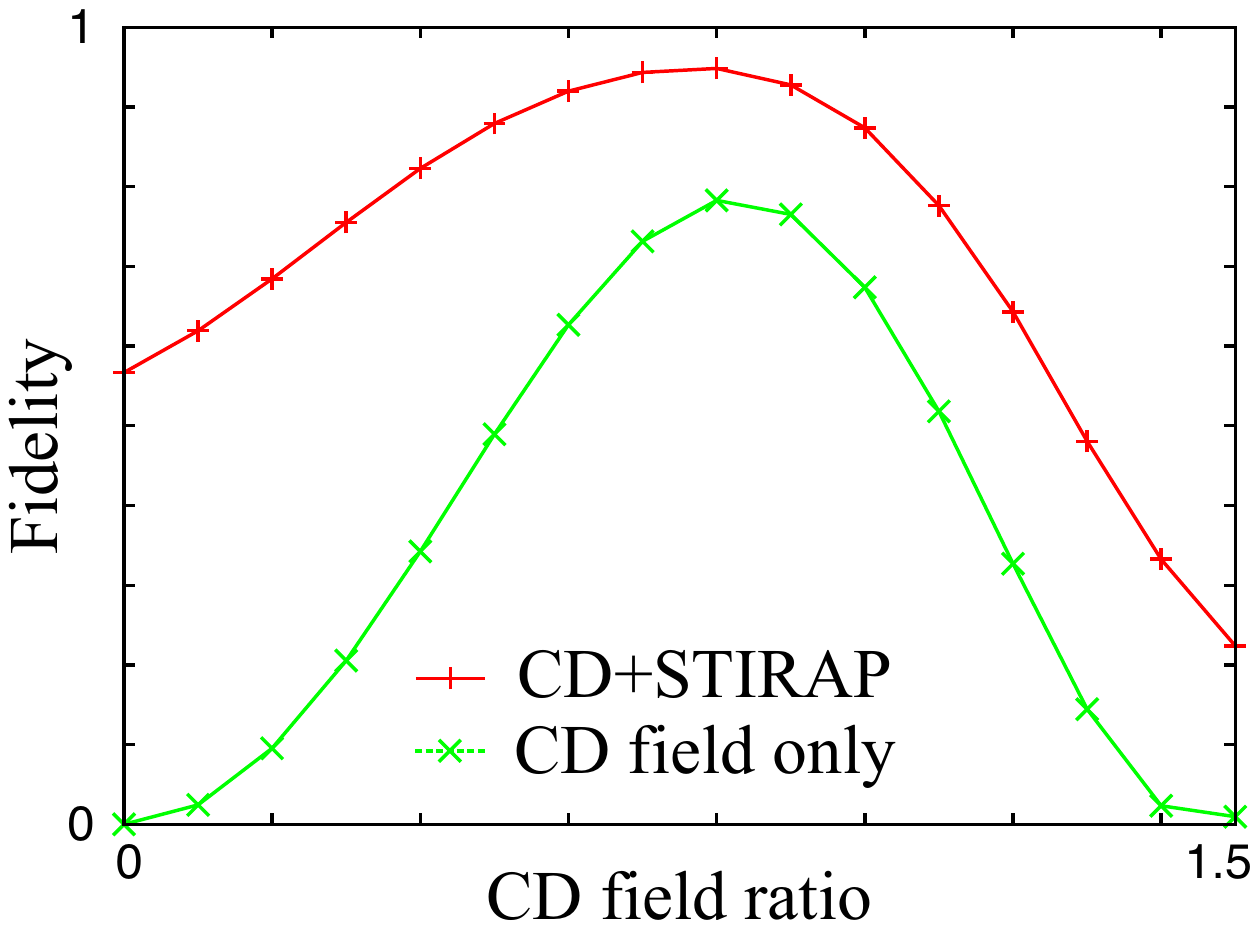}
\end{center}
\caption{ Comparison of the fidelity of STIRAP + CDF control and CDF alone control for various $\lambda$ with FWHM$=212.5$ ps and $T_{2p}-T_{2S} = \mbox{FWHM} / (2\sqrt{\ln 2})$}.
\label{p_cd_cdonly_HCN}
\end{figure}

To study the stability of the efficiency of the population transfer to the variation of the amplitude of the CDF we consider the total driving field in the form displayed in \ref{E_lambda}.  The  $\lambda-$dependence of the fidelity is shown in \ref{p_cd_cdonly_HCN} for STIRAP + CDF control and CDF control.  The combination of STIRAP and CDF generates higher fidelity than either does individually over a wide range of $\lambda$.  It is seen that the fidelity degraded by interference with background states when only STIRAP control is used is restored by the CDF for $\lambda\le 1.2$.

\begin{figure}[h!]
\begin{center}
\includegraphics[width=8cm]{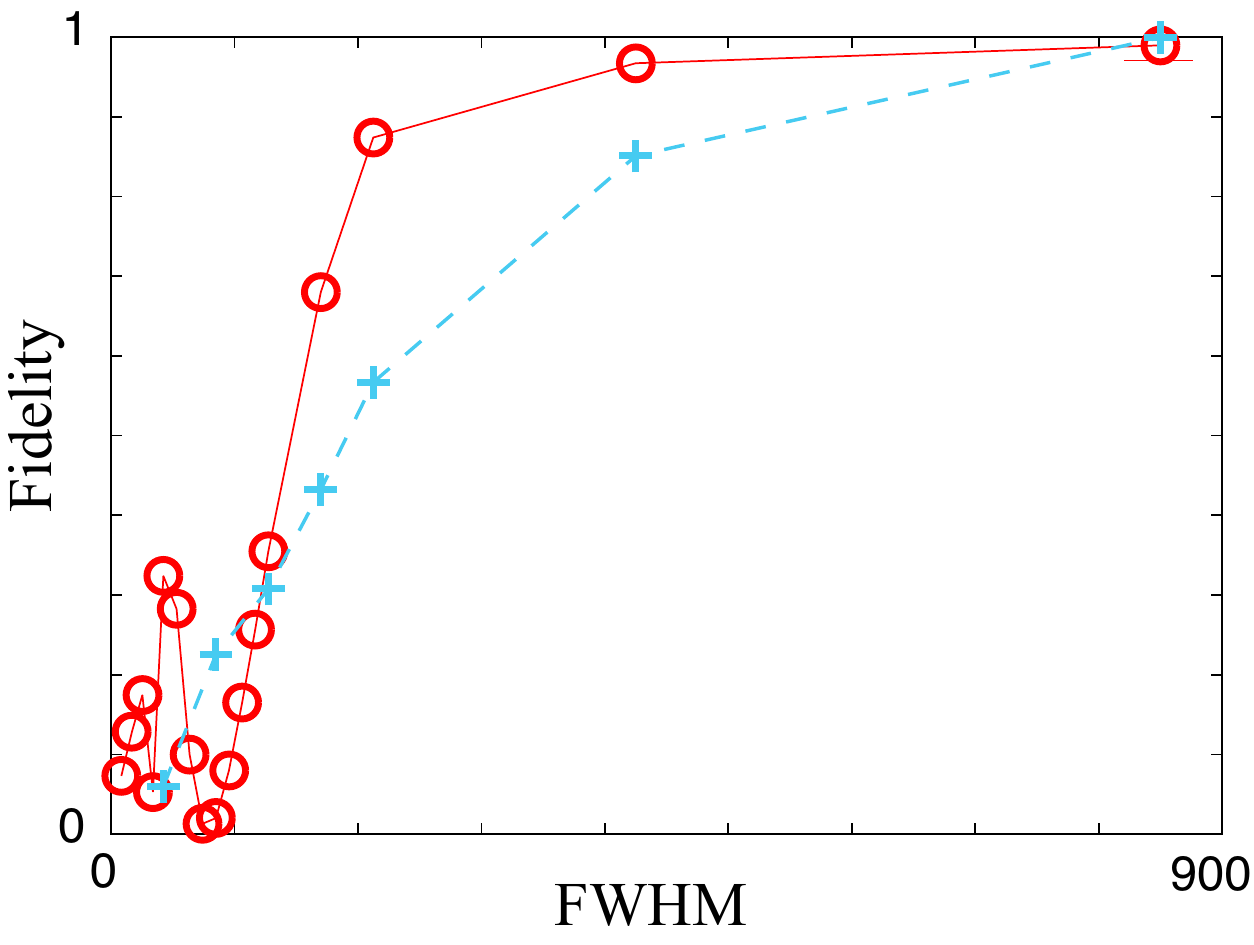}
\end{center}
\caption{ FWHM-dependence of fidelity for STIRAP + CDF control (open
circles) and for STIRAP (crosses) with $T_{2p}-T_{2S} = \mbox{FWHM} / (2\sqrt{\ln 2})$. Fidelity is defined as population for $|5\rangle$ at the final time. The unit of time is ps.}
\label{p_com}
\end{figure}

Consider now the FWHM-dependence of the fidelity. The FWHM-dependence of the fidelity is shown in \ref{p_com}, where open circles and crosses correspond to STIRAP + CDF control and STIRAP control, respectively.  The fidelity of both control methods is less than 1 because of the existence of the background states, and it increases with increasing FWHM for FWHM $>$ 125 ps.  The CDF generates a larger fidelity than does simple STIRAP for FWHM $>$ 125 ps.  The drop of fidelity when FWHM $<$ 125 ps for the STIRAP + CDF control is due to the large intensity of the CDF, which is proportional to 1$/$FWHM.  The results displayed in \ref{p_com} show that the STIRAP + CDF control is more robust to the influence of background states if the amplitude of the CDF is not too large.

\begin{figure}[h!]
\begin{center}
\includegraphics[width=8cm]{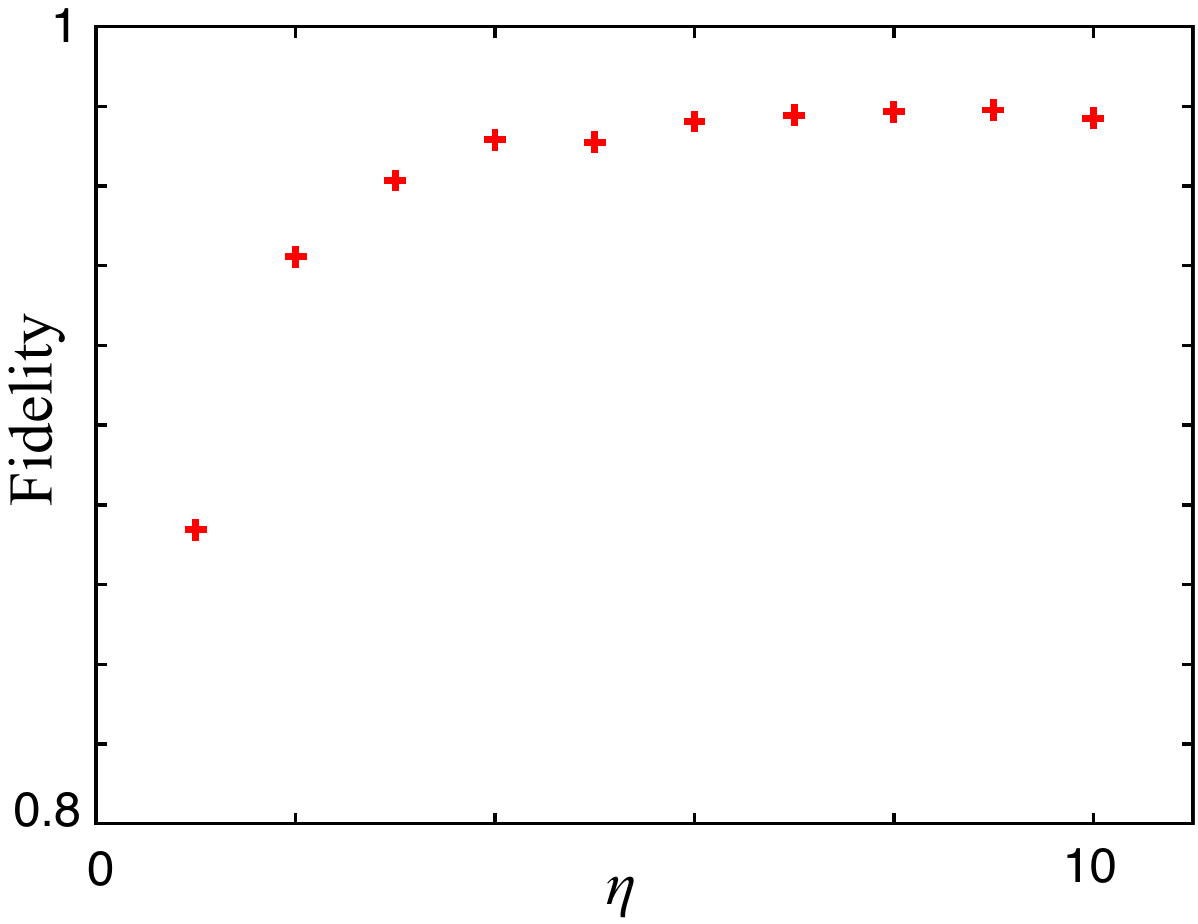}
\end{center}
\caption{ Dependence of the fidelity on $\eta$ for FWHM= 212.5 ps.}
\label{p_com_dT}
\end{figure}

\begin{figure}[h!]
\begin{center}
\includegraphics[width=8cm]{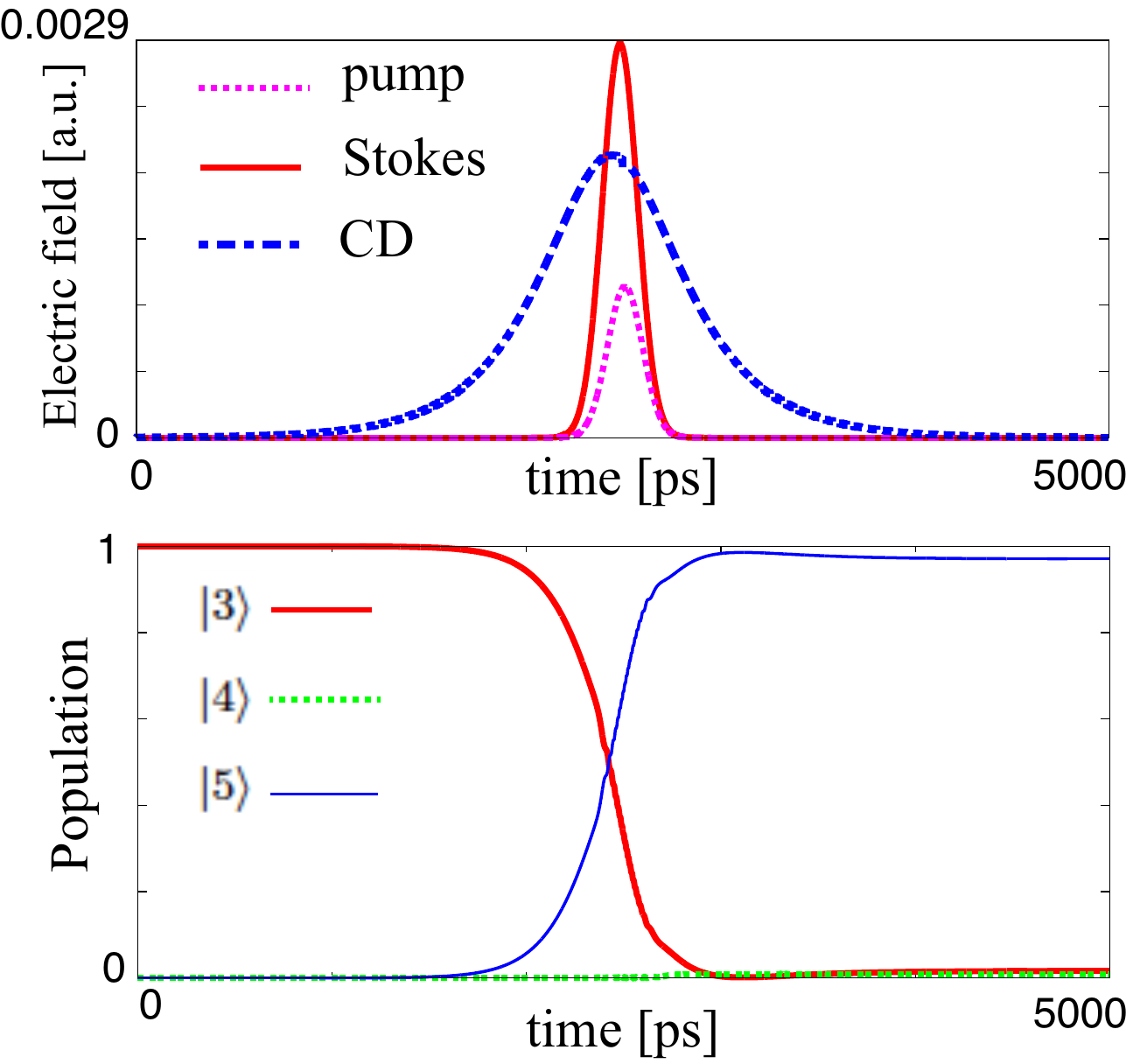}
\end{center}
\caption{Time-dependence of the pump, Stokes and CDFs for FWHM= 212.5 ps and  $\eta=5$ (upper panel). Time-dependences of the population of with STIRAP + CDF control for FWHM = 212.5 ps and $\eta=5$ (lower panel).}
\label{E_p_FWHM212.5_dT5}
\end{figure}

Although ordinary STIRAP becomes less effective if the interval between the Stokes and the pump lasers is small, higher fidelity is obtained by using shorter $T_{2p}-T_{2S}$ in the STIRAP + CDF control process.  The width of the counter-diabatic pulse field becomes large for small $T_{2p}-T_{2S}$.  On the other hand, the peak intensity of the CDF becomes small in that case because it is proportional to $T_{2p}-T_{2S}$.  If the intensity of the CDF is small, unwanted disturbance is diminished.  So as to study the efficiency of the population transfer for various values of $T_{2p}-T_{2S}$ we represent the interval in the form
\begin{eqnarray}
T_{2p}-T_{2S}=\frac{\mbox{FWHM}}{2\eta\sqrt{\ln 2}}.
\end{eqnarray}
Note that $\eta$ defines the interval between the Stokes and pump pulses.  The $\eta$-dependence of the fidelity for FWHM= 212.5 ps is shown in \ref{p_com_dT}.  The time-dependences of the STIRAP + CDFs are shown for $\eta=5$ in the upper panel of \ref{E_p_FWHM212.5_dT5}, and the time-dependences of the populations of $|3\rangle$, $|4\rangle$ and $|5\rangle$ in the STIRAP + CDF control are exhibited in the lower panel in \ref{E_p_FWHM212.5_dT5}.  The time-dependences of the populations of $|3\rangle$, $|4\rangle$ and $|5\rangle$ are similar to those found for adiabatic population transfer in a manifold without the background states.

\section{Discussion}
We have examined an assisted adiabatic transfer scheme designed to control population transfer in a quantum many-body system.  That control is achieved by active manipulation with external fields that work cooperatively with coherence and interference effects embedded in the system quantum dynamics.  Although it is only one of the many schemes that have been proposed to induce complete transfer of population from an arbitrary initial state to a selected target state of a system, our use of STIRAP + CDF control illuminates the generic character of the problems to be solved and the trade-offs that must be made to apply theory to real experimental situations.  In particular, we have examined the role of coupling of the subset of active states, i.e. those used to define the dynamics of interest, with background states.  The results of this examination lead us to emphasize two aspects of the use of isolated subsets of states or reduced Hamiltonians in the development of a control protocol.  

First, the system to be controlled typically has a complicated and dense spectrum of states.  The identification of the initial and target states and relevant intermediate states selects a subset that is embedded in the remaining manifold of states of the system, and the latter form a background that is coupled to the former.  Then, any control protocol must be designed to minimize interference between the background states and excitations within the selected subset of states.  In general, one cannot expect the molecular dynamics of the subset of states to be reducible to the dynamics of an isolated two-level or three-level model spectrum of states, which requires that the consequences of the interactions between the subset states and the background states be managed by the control protocol.  We have shown for the case of vibrational population transfer in SCCl$_2$ that such management can be achieved within a STIRAP + CDF protocol.  Moreover, we have shown that the exact CDF need not be used to obtain very high efficiency of population transfer; an approximate CDF, based on a plausible physical approximation, works very well.  This result is in accord with the findings of Etinski, Uiberacker and Jakubetz \cite{Etinski}.

Second, the counter-diabatic field aided vibrational energy transfer we have discussed modifies the STIRAP process.  Broadly put, the goal of an assisted adiabatic process is the complete transfer of population from an initial state to a selected target state along the adiabatic path.  The STIRAP process, when applicable, achieves the same goal and can be thought of as a special case of an assisted adiabatic process.  With this interpretation in mind, it is worth noting that the STIRAP process has been generalized to deal with population transfer in manifolds that are more complex than the original $|1\rangle$,$|2\rangle$,$|3\rangle$ tripod of states.  For example, Gong and Rice \cite{Gong} have shown how to design a control field that generates complete population transfer from an initial state, via $N$ non-degenerate intermediate states, to one arbitrarily chosen member of $M (M \le N)$ degenerate states.  The full control field exploits a $(M+N-1)$-node null adiabatic state that is created by designing the relative phases and amplitudes of the component fields that together make up the full field.  This solution to the population transfer problem, which can be considered another special case of the counter-diabatic protocol, is universal in the sense that it does not depend on the exact number of the unwanted degenerate states or their properties. Furthermore, this solution suggests that population transfer in a class of multilevel quantum systems with degenerate states can be completely controllable even under strong constraints, e.g., never populating a Hilbert subspace that is only a few dimensions smaller than the whole Hilbert space.  However, this scheme for control of population transfer in a  $(M+N+1)$-level system requires $2N$ pulsed fields and is experimentally unrealistic for, say, $M \ge 3$.  And, if the N intermediate states are a subset of the states in the full manifold it is likely that the use of $2N$ fields will generate interferences with the background states that will compromise the population transfer.

Theoretical efforts devoted to developing control protocols that focus attention on the dynamics of population transfer in isolated few-level discrete systems resemble, to a certain extent, proving trigonometric identities.  Accepting that interference can be exploited in a very large number of ways, involving many different choices of combination of state-to-state pathways, it is possible to develop an almost unlimited number of schemes that in principle generate complete transfer of population from an initial state to a target state for an isolated manifold of discrete states.  However, the complexity of the control protocol will typically preclude experimental exploitation.  And, more important, for a real system the embedding of the states of interest in an environment of other states will necessarily generate interaction that must be accounted for.  The thrust of our argument is simple: Protocols that control the quantum dynamics within a subset of states in a larger manifold of states must be designed to include minimization of the interference between the background states and excitations within the selected subset of states.

\begin{acknowledgement}
S.M. thanks the Grants-in-Aid for Centric Research of Japan Society for Promotion of Science and the JSPS Postdoctoral Fellowships for Research Abroad program for financial support.
\end{acknowledgement}

\appendix{}
\onecolumn
\section{Appendix: Tables of the Energies and the Transition Dipole Moments}
\begin{table*}[h!]
  \caption*{Table A1: Energies of vibrational levels of SCCl$_2$ (From Ref. 
\cite{Kurk2})}
  \begin{tabular}{lccccr}
    \hline
     & energy & & energy & & energy \\
    state & (cm$^{-1}$) & state & (cm$^{-1}$) & state & (cm$^{-1}$) \\
    \hline
    $|1\rangle$ & 4338.6025 & $|8\rangle$ & 5565.4150 & $|15\rangle$ & 5610.4298\\
    $|2\rangle$ & 4543.9591 & $|9\rangle$ & 5658.1828 & $|16\rangle$ & 5947.5765\\
    $|3\rangle$ & 5697.7237 & $|10\rangle$ & 4517.4050 & $|17\rangle$ & 6153.0848\\
    $|4\rangle$ & 5789.3813 & $|11\rangle$ & 4723.0281 & $|18\rangle$ & 6105.2664\\
    $|5\rangle^a$ & 5197.0874 & $|12\rangle$ & 4854.4331 & $|19\rangle$ & 5533.4070\\
    $|5\rangle^b$ & 3171.5275 & $|13\rangle$ & 5327.3689 & $|20\rangle$ & 5485.5693\\
    $|6\rangle$ & 5651.5617 & $|14\rangle$ & 5452.3746 & $|21\rangle$ & 5491.2752\\
    $|7\rangle$ & 5799.4153 &  & &  & \\
  \end{tabular}
\label{table_energy}
\end{table*}

\begin{table*}[h!]
  \caption*{Table A2: Transitions and the transition dipole moments (TDMs) in SCCl$_2$
(From Ref. \cite{Kurk2})}
  \begin{tabular}{l|c|c|c|c|c|c|c}
    \hline
    $|i\rangle\rightarrow|j\rangle$ & TDM & $|i\rangle\rightarrow|j\rangle$ & TDM & $|i\rangle\rightarrow|j\rangle$
      & TDM  & $|i\rangle\rightarrow|j\rangle$ & TDM    \\
    \hline
    $1\rightarrow 2$ & 0.09805 & $2\rightarrow 5^a$ & 0.2283 & $3\rightarrow 14$ & 0.4146 & $5^a\rightarrow 9$ & 0.4138 \\
    $2\rightarrow 3$ & 0.2062 & $2\rightarrow 5^b$ & 0.02062 & $3\rightarrow 15$ & 0.10125 & $5^a\rightarrow 10$ & 0.02298  \\
    $2\rightarrow 4$ & 0.2283 & $2\rightarrow 6$ & 0.1039 & $3\rightarrow 16$ & 0.23235 & $5^a\rightarrow 11$ & 0.03331  \\
    $3\rightarrow 5^a$ & 0.09805 & $2\rightarrow 7$ & 0.2283 & $3\rightarrow 17$ & 0.2090 & $5^a\rightarrow 12$ & 0.2298  \\
    $4\rightarrow 5^a$ & 0.0022616 & $2\rightarrow 8$ & 0.009805 & $3\rightarrow 18$ & 0.1002 & $5^a\rightarrow 13$ & 0.2078  \\
    $3\rightarrow 5^b$ & 0.002077 & $2\rightarrow 9$ & 0.3451 & $3\rightarrow 19$ & 0.2062 & $5^a\rightarrow 14$ & 0.4028  \\
    $4\rightarrow 5^b$ & 0.1015 & $2\rightarrow 10$ & 0.2283 & $3\rightarrow 20$ & 0.04025 & $5^a\rightarrow 15$ & 0.014765  \\
    $1\rightarrow 3$ & 0.2283 & $2\rightarrow 11$ & 0.2080 & $3\rightarrow 21$ & 0.1015 & $5^a\rightarrow 16$ & 0.21065  \\
    $1\rightarrow 4$ & 0.2062 & $2\rightarrow 12$ & 0.02616 & $4\rightarrow 6$ & 0.2090 & $5^a\rightarrow 17$ & 0.2309  \\
    $1\rightarrow 5^a$ & 0.2062 & $2\rightarrow 13$ & 0.1015 & $4\rightarrow 7$ & 0.002977 & $5^a\rightarrow 18$ & 0.02062  \\
    $1\rightarrow 5^b$ & 0.1002 & $2\rightarrow 14$ & 0.3597 & $4\rightarrow 8$ & 0.2062 & $5^a\rightarrow 19$ & 0.2283  \\
    $1\rightarrow 6$ & 0.03448 & $2\rightarrow 15$ & 0.22875 & $4\rightarrow 9$ & 0.4138 & $5^a\rightarrow 20$ & 0.04025  \\
    $1\rightarrow 7$ & 0.2061 & $2\rightarrow 16$ & 0.1072 & $4\rightarrow 10$ & 0.02283 & $5^a\rightarrow 21$ & 0.02616  \\
    $1\rightarrow 8$ & 0.0009805 & $2\rightarrow 17$ & 0.03448 & $4\rightarrow 11$ & 0.03332 & $5^b\rightarrow 6$ & 0.1060  \\
    $1\rightarrow 9$ & 0.3588 & $2\rightarrow 18$ & 0.2283 & $4\rightarrow 12$ & 0.2298 & $5^b\rightarrow 7$ & 0.09808  \\
    $1\rightarrow 10$ & 0.2064 & $2\rightarrow 19$ & 0.002536 & $4\rightarrow 13$ & 0.2078 & $5^b\rightarrow 8$ & 0.02283  \\
    $1\rightarrow 11$ & 0.2078 & $2\rightarrow 20$ & 0.2309 & $4\rightarrow 14$ & 0.4028 & $5^b\rightarrow 9$ & 0.3457  \\
    $1\rightarrow 12$ & 0.1015 & $2\rightarrow 21$ & 0.2298 & $4\rightarrow 15$ & 0.02540 & $5^b\rightarrow 10$ & 0.2298  \\
    $1\rightarrow 13$ & 0.02628 & $3\rightarrow 4$ & 0.09808 & $4\rightarrow 16$ & 0.21065 & $5^b\rightarrow 11$ & 0.2080  \\
    $1\rightarrow 14$ & 0.3461 & $3\rightarrow 6$ & 0.2308 & $4\rightarrow 17$ & 0.2309 & $5^b\rightarrow 12$ & 0.02078  \\
    $1\rightarrow 15$ & 0.2067 & $3\rightarrow 7$ & 0.09806 & $4\rightarrow 18$ & 0.02629 & $5^b\rightarrow 13$ & 0.1003  \\
    $1\rightarrow 16$ & 0.04328 & $3\rightarrow 8$ & 0.2064 & $4\rightarrow 19$ & 0.025355 & $5^b\rightarrow 14$ & 0.3594  \\
    $1\rightarrow 17$ & 0.1039 & $3\rightarrow 9$ & 0.4020 & $4\rightarrow 20$ & 0.04034 & $5^b\rightarrow 15$ & 0.2288  \\
    $1\rightarrow 18$ & 0.2062 & $3\rightarrow 10$ & 0.1002 & $4\rightarrow 21$ & 0.02616 & $5^b\rightarrow 16$ & 0.1060  \\
    $1\rightarrow 19$ & 0.09808 & $3\rightarrow 11$ & 0.03472 & $5^a\rightarrow 6$ & 0.2090 & $5^b\rightarrow 17$ & 0.04328  \\
    $1\rightarrow 20$ & 0.2090 & $3\rightarrow 12$ & 0.2078 & $5^a\rightarrow 7$ & 0.001421 & $5^b\rightarrow 18$ & 0.2283  \\
    $1\rightarrow 21$ & 0.3331 & $3\rightarrow 13$ & 0.2298 & $5^a\rightarrow 8$ & 0.2062 & $5^b\rightarrow 19$ & 0.02077  \\
  \end{tabular}
\label{table_TDM}
\end{table*}

\begin{table*}[h!]
  \caption*{Table A3: Transitions and the transition dipole moments (TDMs) in SCCl$_2$
(From Ref. \cite{Kurk2})}
  \begin{tabular}{l|c|c|c|c|c|c|r}
    \hline
    $|i\rangle\rightarrow|j\rangle$ & TDM & $|i\rangle\rightarrow|j\rangle$ & TDM & $|i\rangle\rightarrow|j\rangle$
      & TDM  & $|i\rangle\rightarrow|j\rangle$ & TDM   \\
    \hline
$5^b\rightarrow 20$ & 0.2309 &   $8\rightarrow 9$ & 0.3452 &   $10\rightarrow 17$ & 0.2309    & $13\rightarrow 21$ & 0.2062 \\
$5^b\rightarrow 21$ & 0.09810 & $8\rightarrow 10$ & 0.2283 & $10\rightarrow 18$ & 0.01015 & $14\rightarrow 15$ & 0.21065 \\
$6\rightarrow 7$ & 0.2090 &         $8\rightarrow 11$ & 0.2078 & $10\rightarrow 19$ & 0.2283     & $14\rightarrow 16$ & 0.3451 \\
$6\rightarrow 8$ & 0.3450 &         $8\rightarrow 12$ & 0.02835 &$10\rightarrow 20$ & 0.03603 & $14\rightarrow 17$ & 0.3597 \\
$6\rightarrow 9$ & 0.3588 &         $8\rightarrow 13$ & 0.02630 &$10\rightarrow 21$ & 0.03472 & $14\rightarrow 18$ & 0.2083 \\
$6\rightarrow 10$ & 0.2093 &       $8\rightarrow 14$ & 0.3461 & $11\rightarrow 12$ & 0.2064    & $14\rightarrow 19$ & 0.3597 \\
$6\rightarrow 11$ & 0.21065 &     $8\rightarrow 15$ & 0.2067 &  $11\rightarrow 13$ & 0.2062   & $14\rightarrow 20$ & 0.40285 \\
$6\rightarrow 12$ & 0.1072 &       $8\rightarrow 16$ & 0.04329 & $11\rightarrow 14$ & 0.4020  & $14\rightarrow 21$ & 0.4020 \\
$6\rightarrow 13$ & 0.02628 &     $8\rightarrow 17$ & 0.03585 & $11\rightarrow 15$ & 0.02998 &  $15\rightarrow 16$ & 0.4028 \\
$6\rightarrow 14$ & 0.3461 &       $8\rightarrow 18$ & 0.2062 & $11\rightarrow 16$ & 0.2090    & $15\rightarrow 17$ & 0.4135 \\
$6\rightarrow 15$ & 0.4020 &       $8\rightarrow 19$ & 0.01013 & $11\rightarrow 17$ & 0.2109  & $15\rightarrow 18$ & 0.3451 \\
$6\rightarrow 16$ & 0.02616 &    $8\rightarrow 20$ & 0.2093 & $11\rightarrow 18$ & 0.02618   & $15\rightarrow 19$ & 0.2313 \\
$6\rightarrow 17$ & 0.09805 &     $8\rightarrow 21$ & 0.2078 &  $11\rightarrow 19$ & 0.2081    &     $15\rightarrow 20$ & 0.3451 \\
$6\rightarrow 18$ & 0.2090 &       $9\rightarrow 10$ & 0.2283 &  $11\rightarrow 20$ & 0.0360    & $15\rightarrow 21$ & 0.03637 \\
$6\rightarrow 19$ & 0.1040 &       $9\rightarrow 11$ & 0.4030 &$11\rightarrow 21$ & 0.02063    & $16\rightarrow 17$ & 0.1015\\
$6\rightarrow 20$ & 0.2062 &      $9\rightarrow 12$ & 0.3461&$12\rightarrow 13$ & 0.9808        & $16\rightarrow 18$ & 0.4026\\
$6\rightarrow 21$ & 0.21065 &    $9\rightarrow 13$ & 0.4440 & $12\rightarrow 14$ & 0.3588     &  $16\rightarrow 19$ & 0.1072\\
$7\rightarrow 8$ & 0.2062 &        $9\rightarrow 14$ & 0.1015 &  $12\rightarrow 15$ & 0.2302    & $16\rightarrow 20$ & 0.2078\\
$7\rightarrow 9$ & 0.4138 &        $9\rightarrow 15$ & 0.2309 & $12\rightarrow 16$ & 0.1039     & $17\rightarrow 21$ & 0.2090\\
$7\rightarrow 10$ & 0.02302 &    $9\rightarrow 16$ & 0.3597 & $12\rightarrow 17$ & 0.04328  & $17\rightarrow 18$ & 0.2309\\
$7\rightarrow 11$ & 0.03335 &   $9\rightarrow 17$ & 0.3451 & $12\rightarrow 18$ & 0.2298     & $17\rightarrow 19$ & 0.03457\\
$7\rightarrow 12$ & 0.2298 &      $9\rightarrow 18$ & 0.2288  &$12\rightarrow 19$ & 0.02628  & $17\rightarrow 20$ & 0.2283\\
$7\rightarrow 13$ & 0.2078 &      $9\rightarrow 19$ & 0.3451  &$12\rightarrow 20$ & 0.2324    &$17\rightarrow 21$ & 0.23235\\
$7\rightarrow 14$ & 0.4028 &      $9\rightarrow 20$ & 0.4138  &  $12\rightarrow 21$ & 0.2283  & $18\rightarrow 19$ & 0.2283\\
$7\rightarrow 15$ & 0.02527 &    $9\rightarrow 21$ & 0.4146  & $13\rightarrow 14$ & 0.3451   & $18\rightarrow 20$ & 0.03457\\
$7\rightarrow 16$ & 0.2107 &      $10\rightarrow 11$ & 0.1015 &   $13\rightarrow 15$ & 0.2083 & $18\rightarrow 21$ & 0.03330\\
$7\rightarrow 17$ & 0.2309 &      $10\rightarrow 12$ & 0.2298 & $13\rightarrow 16$ & 0.03457 & $19\rightarrow 20$ & 0.2309\\
$7\rightarrow 18$ & 0.02066 &    $10\rightarrow 13$ & 0.2081 & $13\rightarrow 17$ & 0.1072   & $19\rightarrow 21$ & 0.2298\\
$7\rightarrow 19$ & 0.2283 &      $10\rightarrow 14$ & 0.4030 &     $13\rightarrow 18$ & 0.2078 & $20\rightarrow 21$ & 0.04800\\
$7\rightarrow 20$ & 0.04028 &    $10\rightarrow 15$ & 0.03153 &$13\rightarrow 19$ & 0.1015   & &\\
$7\rightarrow 21$ & 0.02620 &    $10\rightarrow 16$ & 0.210      &$13\rightarrow 20$ & 0.21065 & &
  \end{tabular}
\label{table_TDM2}
\end{table*}

\begin{table*}[htb]
  \caption*{Table A4: Transitions and the transition dipole moments associated with
  the levels considered (From Ref.\cite{Kurk1})}
  \begin{tabular}{lr}
    \hline
      Transition & Transition dipole moment (a.u.)  \\
     \hline \hline
     (0,0,0) of HCN $\rightarrow$ (1,0,1) of HCN \hspace{1cm}& 0.0044\\
     (0,0,0) of HCN $\rightarrow$ (2,0,1) of HCN \hspace{1cm}& 0.0006892\\
     (0,0,0) of HCN $\rightarrow$ (3,0,1) of HCN \hspace{1cm}& 0.0001860\\
     (0,0,0) of HCN $\rightarrow$ (3,0,2) of HCN \hspace{1cm}& 0.00002918\\
     (0,0,0) of HCN $\rightarrow$ (5,0,1)  \hspace{1cm}& 0.00003225\\
     (0,0,0) of HCN $\rightarrow$ (6,0,2)  \hspace{1cm}& 0.000007044\\
     (1,0,1) of HCN $\rightarrow$ (2,0,1)  \hspace{1cm}& 0.087715\\
     (1,0,1) of HCN $\rightarrow$ (3,0,1) of HCN \hspace{1cm}& 0.00739\\
     (1,0,1) of HCN $\rightarrow$ (3,0,2) of HCN \hspace{1cm}& 0.0007\\
     (1,0,1) of HCN $\rightarrow$ (5,0,1) of HCN \hspace{1cm}& 0.000168\\
     (1,0,1) of HCN $\rightarrow$ (6,0,2)  \hspace{1cm}& 0.0000316\\
     (2,0,1) of HCN $\rightarrow$ (3,0,1)  \hspace{1cm}& 0.08917\\
     (2,0,1) of HCN $\rightarrow$ (3,0,2) of HCN \hspace{1cm}& 0.00452\\
     (2,0,1) of HCN $\rightarrow$ (5,0,1) of HCN \hspace{1cm}& 0.001069\\
     (2,0,1) of HCN $\rightarrow$ (6,0,2)  \hspace{1cm}& 0.0000607\\
     (3,0,1) of HCN $\rightarrow$ (3,0,2) of HCN \hspace{1cm}& 0.004871\\
     (3,0,1) of HCN $\rightarrow$ (5,0,1) of HCN \hspace{1cm}& 0.007626\\
     (3,0,1) of HCN $\rightarrow$ (6,0,2)  \hspace{1cm}& 0.0002267\\
     (3,0,2) of HCN $\rightarrow$ (5,0,1) of HCN \hspace{1cm}& 0.001026\\
     (3,0,2) of HCN $\rightarrow$ (6,0,2)  \hspace{1cm}& 0.0013285\\
     (5,0,1) $\rightarrow$ (6,0,2)  \hspace{1cm}& 0.007314\\
     (5,0,1) $\rightarrow$ (2,0,1) of HNC \hspace{1cm}& 0.003054\\
     (5,0,1) $\rightarrow$ (2,0,0) of HNC \hspace{1cm}& 0.0002726\\
     (5,0,1) $\rightarrow$ (1,0,1) of HNC \hspace{1cm}& 0.0000764\\
     (5,0,1) $\rightarrow$ (0,0,0) of HNC \hspace{1cm}& 0.00003676\\
     (6,0,2) $\rightarrow$ (2,0,1) of HNC \hspace{1cm}& 0.001163\\
     (6,0,2) $\rightarrow$ (2,0,0) of HNC \hspace{1cm}& 0.00003136\\
     (6,0,2) $\rightarrow$ (1,0,1) of HNC \hspace{1cm}& 0.00004371\\
     (6,0,2) $\rightarrow$ (0,0,0) of HNC \hspace{1cm}& 0.0000183\\
     (2,0,1) $\rightarrow$ (2,0,0) of HNC \hspace{1cm}& 0.1068\\
     (2,0,1) $\rightarrow$ (1,0,1) of HNC \hspace{1cm}& 0.16174\\
     (2,0,1) $\rightarrow$ (0,0,0) of HNC \hspace{1cm}& 0.0009863\\
     (2,0,0) $\rightarrow$ (1,0,1) of HNC \hspace{1cm}& 0.01962\\
     (2,0,0) $\rightarrow$ (0,0,0) of HNC \hspace{1cm}& 0.016485\\
     (1,0,1) $\rightarrow$ (0,0,0) of HNC \hspace{1cm}& 0.0100926\\\hline     
  \end{tabular}
\label{table_d_1}
\end{table*}


\begin{table*}[htb]
  \caption*{Table A5: The energies of the vibrational states (From Ref. \cite{Kurk1})}
  \begin{tabular}{|l|c|c|}
    \hline
      ($\nu_1$,$\nu_2$,$\nu_3$) & $|i\rangle$ & energy [1/cm]\\
     \hline \hline
     (0,0,0) HCN & $|1\rangle$ & 0.0\\
     (2,0,1) HCN & $|2\rangle$ & 8585.87\\
     (5,0,1) & $|3\rangle$ & 17574.40\\
     (2,0,1) HNC & $|4\rangle$ & 14154.30\\
     (0,0,0) HNC & $|5\rangle$ & 5023.20\\
     (1,0,1) HCN & $|6\rangle$ & 5393.70\\
     (3,0,1) HCN & $|7\rangle$ & 11674.45\\
     (3,0,2) HCN & $|8\rangle$ & 13702.24\\
     (6,0,2) & $|9\rangle$ & 19528.57\\
     (2,0,0) HNC & $|10\rangle$ & 12139.90\\
     (1,0,1) HNC & $|11\rangle$ & 10651.90\\

\hline
  \end{tabular}
\label{table_e}
\end{table*}


\clearpage

\twocolumn

\end{document}